\documentclass[]{aa} 
\usepackage{float}
\usepackage{graphicx}
\usepackage{txfonts}
\usepackage[utf8]{inputenc}
\usepackage[T1]{fontenc}
\usepackage{amsfonts}
\usepackage{natbib}
\usepackage{pdflscape}
\usepackage{longtable}
\usepackage{float}
\usepackage{lscape}
\usepackage{nicefrac}
\usepackage{multirow}
\usepackage{multicol}
\usepackage[urlcolor=cyan, colorlinks=true, citecolor=blue, linkcolor=blue]{hyperref}

\newcommand{\mysp}{}\def\mysp/{} 
\newcommand{\iras}{IRAS\,22147$+$5948}

\title{A census of young stellar objects in two line-of-sight star-forming regions toward \iras{} in the outer Galaxy}

\begin{document}

\author{Agata Karska\inst{1}, Maciej Koprowski\inst{1}, Aleksandra Solarz\inst{2}, Ryszard Szczerba\inst{3}, Marta Sewi{\l}o\inst{4,5,6}, Natasza Si\'odmiak\inst{3}, Davide Elia\inst{7}, Marcin Gawroński\inst{1}, Konrad Grzesiak\inst{1}, Bosco H. K. Yung\inst{3}, William J. Fischer\inst{8}, Lars E. Kristensen\inst{9}}  
        
\institute{$^{1}$ Institute of Astronomy, Faculty of Physics, Astronomy and Informatics, Nicolaus Copernicus University, ul. Grudziądzka 5, 87-100 Toruń, Poland\\ 
$^{2}$ European Southern Observatory, Alonso de Cordova 3107, Casilla 19001, Santiago, Chile \\
$^{3}$ Nicolaus Copernicus Astronomical Center, ul. Rabiańska 8, 87-100 Toruń, Poland \\ 
$^{4}$ Exoplanets and Stellar Astrophysics Laboratory, NASA Goddard Space Flight Center, Greenbelt, MD 20771, USA\\
$^{5}$ Department of Astronomy, University of Maryland, College Park, MD 20742, USA \\
$^{6}$ Center for Research and Exploration in Space Science and Technology, NASA Goddard Space Flight Center, Greenbelt, MD 20771
$^{7}$ Istituto di Astrofisica e Planetologia Spaziali-INAF, Via Fosso del Cavaliere 100, I-00133 Roma, Italy \\
$^{8}$ Space Telescope Science Institute, 3700 San Martin Dr., Baltimore, MD 21218, USA\\
$^{9}$ Centre for Star and Planet Formation, Niels Bohr Institute, University of Copenhagen, \O ster Voldgade 5-7, DK-1350 Copenhagen K, Denmark\\
}

\date{Received June 30, 2021; accepted May 2, 2022}
\titlerunning{YSOs in the IRAS 22147 region}
\authorrunning{A.~Karska et al. 2021}

\abstract
{Star formation in the outer Galaxy, namely, outside of the Solar circle, has not been  extensively studied in part due to the low CO brightness of the molecular clouds linked with the negative metallicity gradient. Recent infrared surveys provide an overview of dust emission in large sections of the Galaxy, but they suffer from cloud confusion and poor spatial resolution at far-infrared wavelengths.}
{We aim to develop a methodology to identify and classify young stellar objects (YSOs) in star-forming regions in the outer Galaxy and use it to resolve a long-standing disparity in terms of the distance and evolutionary status of \iras{}.}
{We used a support vector machine learning algorithm to complement standard color-color and color-magnitude diagrams in our search for YSOs in the IRAS 22147 region, based on publicly available data from the Spitzer Mapping of the Outer Galaxy survey.
 The agglomerative hierarchical clustering algorithm was used to identify clusters. Then the physical properties of individual YSOs were calculated. The distances were determined using CO 1-0 from the Five College Radio Astronomy Observatory survey.}
{We identified 13 Class I and 13 Class II YSO candidates using the color-color diagrams, along with an additional 2 and 21 sources, respectively, using the applied machine learning techniques. 
The spectral energy distributions of 23 sources were modeled with a star and a passive disk, corresponding to Class II objects. The models of three sources include envelopes that are typical for Class I objects. The objects were grouped into two clusters located at a distance of $\sim2.2$ kpc and 5 clusters at $\sim5.6$ kpc. The spatial extent of CO, radio continuum, and dust emission confirms the origin of YSOs in two distinct star-forming regions along a similar line of sight.}
{The outer Galaxy may serve as a unique laboratory for exploring star formation across environments, on the condition that complementary methods and ancillary data are used to properly account for cloud confusion and distance uncertainties.}

\keywords{stars: formation -- ISM: molecules -- ISM: individual objects:
\iras{}}

\maketitle

\section{Introduction} 
\label{sec:intro}
\begin{figure*} 
\centering
\includegraphics[scale=0.39, trim=0cm 0cm 0cm 0cm]{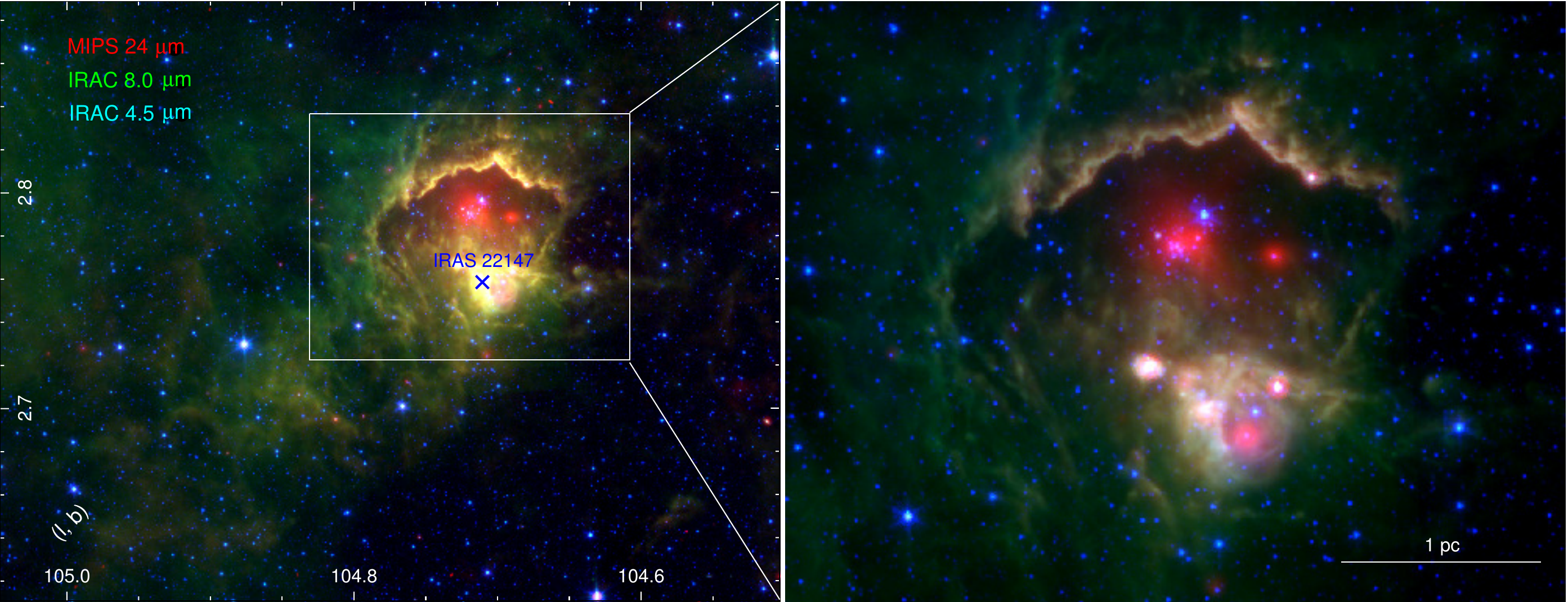}
\caption{Three-color composite image of the IRAS 22147 region combining the \textit{Spitzer}/MIPS 24 $\mu$m (red) and \textit{Spitzer}/IRAC 8.0 $\mu$m (green) and 4.5 $\mu$m (blue) mosaics. The position of \iras{} is indicated with an $\times$ symbol. The right panel shows a zoom-in on the most active region. \label{fig:rgb} }
\end{figure*}

Stars form inside cold and dense cores in molecular clouds. In the Solar neighborhood, about 3000 young stellar objects (YSOs) have been identified in molecular clouds as part of the \textit{Spitzer} Space Telescope Gould Belt survey, providing lifetimes of YSO evolutionary stages and star formation efficiencies in various clouds \citep{dunham2015,kd18}. Similar efforts in the outer Galaxy suffer from low angular resolution, sensitivity, and distance uncertainties, and they are often limited to high-mass objects \citep{ur2008,konig21}. However, due to differences in environmental conditions and a high fraction of atomic-to-molecular gas, the star formation rate at the outer parts of the Galactic disk is expected to be lower \citep{ke12}.

Galactic surveys using CO and its isotopologues have been used to pinpoint the location and determine physical properties of molecular clouds up to about 20 kpc from the Galactic Centre. A negative metallicity gradient in the Galaxy affects the abundances of dust and molecules \citep{sodroski1997}, as well as the overall gas and dust cooling budget \citep{roman2010}. It causes an increase in the CO-to-H$_2$ conversion factor \citep{digel1990,pohl2008,heyer2009} and a decrease in the gas-to-dust ratio with the Galactocentric radius \citep{giannetti2017}. All of these factors likely contribute to the decrease of the mass surface density of molecular clouds in the outer Galaxy \citep[for a review, see ][]{hd15}.

Clearly, a proper characterization of physical conditions and chemistry of star-forming regions in multiple locations is necessary for improving our understanding of the impact of environment on star formation and its efficiency in the outer Galaxy. The initial step, however, is to obtain a census of YSOs in multiple molecular clouds and build a sample for follow-up spectroscopy surveys. Subsequently,  it will be possible to obtain physical and chemical properties of YSOs and link them with the large-scale properties of molecular clouds in the Milky Way. 

Recent infrared sky surveys with the \textit{Spitzer} Space Telescope \citep[\textit{Spitzer}, ][]{werner2004} and the \textit{Herschel} Space Observatory \citep[\textit{Herschel}, ][]{pilbratt2010} provide a complementary picture of ongoing star formation in uncharted regions of the Galaxy. The Spitzer Mapping of the Outer Galaxy (SMOG; PI: S. Carey) survey is specifically aimed to characterize YSOs in sparsely studied star-forming regions around the galactic longitude $l\sim105^{\circ}$ and encompassing Perseus and Outer spiral arms at Galactocentric radii of about 9 and 12 kpc, respectively \citep{carey2008}. A recent study by \citet{Winston2019} identified $\sim4650$ YSO candidates located in 68 clusters and stressed the similarities in star formation properties with the inner Galaxy. 

Similar exploratory studies around $l\sim220^{\circ}$ were performed by combining near-IR photometry with longer wavelength data, which are also sensitive to more deeply-embedded YSOs. \citet{sewilo2019} characterized a very active star formation site CMa$-l224$ consisting of $\sim$290 Class I/II YSOs, using data from GLIMPSE360: Completing the Spitzer Galactic Plane Survey (PI: B. Whitney) and the Herschel infrared Galactic Plane Survey (Hi-GAL; \citealt{molinari2010}, see also \citealt{elia2013}). The region resembles low-mass star-forming regions in the Gould Belt and possesses the largest concentrations of YSOs and their clusters in a 100 deg$^2$ field centered on the Canis Major star-forming region toward $l\sim220^{\circ}$ \citep{fischer16}.

In this paper, we aim to investigate one specific star-forming region located in the $l\sim105^{\circ}$ area, which has been subject to contradictory classifications and distance estimates. Here, we assess to what extent various methods of YSO identification provide consistent results. \iras{} was first detected in CO 1-0 with IRAM 30 m as part of a follow-up observing campaign of the star-forming region candidates among IRAS sources in the outer Galaxy, and assigned a distance of 6.48 kpc \citep[$\varv_\mathrm{LSR}$=-59.05 km s$^{-1}$, ][]{wb89}. In a parallel survey with the Millimeter-Wave Observatory, the distance of 2.8 kpc was determined based on CO 2-1 observations \citep[$\varv_\mathrm{LSR}$=-59.5 km s$^{-1}$, ][]{wilking1989}. The Dominion Radio Astrophysical Observatory survey at $l\sim105^{\circ}$ classified \iras{} as a likely supernova remnant due to a slightly non-thermal spectral index \citep[SNR G104.7+2.8, ][]{joncas1990,green1994}. Subsequent observations from the Canadian Galactic Plane Survey showed that radio spectra are rather flat and this object is most likely an \ion{H}{ii} region \citep{taylor2003,kothes2006}, without, however, any maser detections \citep{molinari1996,Fontani2010}. Infrared observations with 2MASS and WISE provided further evidence that \iras{} is a star-formation site, with stellar surface density enhancement typical for embedded cluster and a blob or shell morphology \citep{kumar2006,lundquist2014}.

Here, we address the most prevalent questions, such as the measurement of the distance to \iras{} and the census of YSOs obtained with standard methods using infrared colors. We also consider whether machine learning techniques utilizing the AllWISE catalog are equipped to provide new YSO identifications. We also ask whether they are grouped in clusters and whether the distribution of gas and dust emission is consistent with the picture of star formation in \iras{}. Finally, we examine whether the spectral energy distributions of YSO candidates are consistent with the models from \citet{robitaille2017}.

The paper is organized as follows. Section\,\ref{sec:data} describes the multi-wavelength data used in this work, including our maser survey. In Section\,\ref{sec:res}, we perform an analysis involving the determination of the kinematic distances to the SMOG sources (Section \ref{sec:dist}), an initial identification of YSOs using color-color cuts (Section \ref{sec:cc}) and machine-learning algorithms (Section \ref{sec:ml}), a comparison to the results from \citealt{Winston2019} (Section \ref{sec:complet}), an investigation of the clustering of YSOs (\ref{sec:clust}) and associations with the interstellar medium (Section \ref{sec:spatial}), finding the best-fit spectral energy distributions with the corresponding physical parameters (\ref{sec:seds}). Section~\ref{sec:disc} provides the discussion of the results in the context of previous studies and Section~\ref{sec:concs} presents our summary and conclusions. 

\section{Data} 
\label{sec:data}

\subsection{\textit{Spitzer} SMOG survey}
\label{sec:smog}
\begin{figure}[ht!]
\includegraphics[width=1.0\columnwidth]{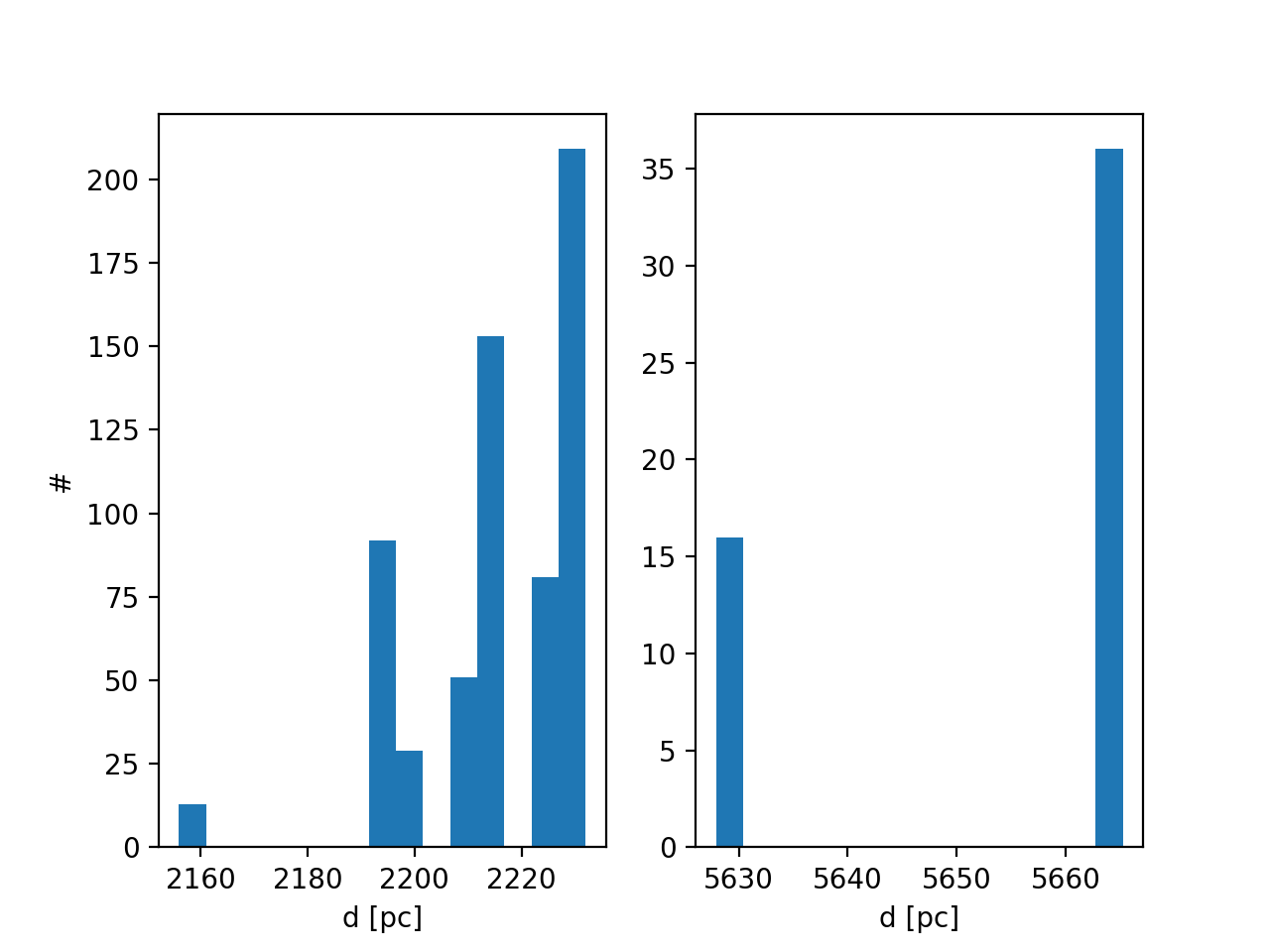} 
\caption{Histogram of distances for sources in the IRAS 22147 region with highly reliable IRAC photometry ($\sigma<0.2$ mag).} 
\label{dist}
\end{figure}

The primary data set used in this study consists of the images and point source catalogs from the SMOG survey \citep{carey2008}. The SMOG survey covers an area of $\sim$21 square degrees in the outer Galactic plane with the Galactic longitude ({\it l}) and Galactic latitude ({\it b}) ranges of roughly (102$^{\circ}$, 109$^{\circ}$) and (0$^{\circ}$, 3$^{\circ}$), respectively, with the Infrared Array Camera (IRAC, 3.6--8.0 $\mu$m; \citealt{fazio2004}) and Multiband Imaging Photometer for Spitzer (MIPS, 24--160 $\mu$m; \citealt{rieke2004}) instruments. In this study, we focus on a small region around \iras{}
($104\rlap.^{\circ}55<l<105^{\circ}$) and ($2\rlap.^{\circ}55<b<2\rlap.^{\circ}95$), which we refer to as the " IRAS 22147 region."

The SMOG IRAC data were processed by the Wisconsin GLIMPSE IRAC pipeline and the data products are available at the Infrared Science Archive (IRSA; \citealt{smogdoc}). The data products include the IRAC point source catalog containing the highest reliability sources and the more complete IRAC point source archive containing sources with less stringent selection criteria than the catalog. Here, we use the IRAC archive, which is more complete, although it requires the identification and removal of unreliable detections during our analysis. The angular resolution of the {\it Spitzer} IRAC 3.6--8.0 $\mu$m observations is $\sim$2$''$.

\textit{Spitzer} data is supplemented with near-IR ({\it JHK$_{s}$}) photometry from the Two Micron All Sky Survey \citep[2MASS, ][]{skrutskie2006}, which is included in the final IRAC catalogs at IRSA. 

Figure 1 shows the three-color composite image of the IRAS 22147 region using IRAC and MIPS mosaics. The morphology of the region reveals an embedded cluster corresponding to the position of \iras{} and an arc of emission to the north. An extended structure
toward south-east corresponds to the position of IRAS 22159+5948. 

 \subsection{AllWISE catalog}
\label{sec:2mass}

The AllWISE program combines the data from the Wide-field Infrared Survey Explorer (WISE; \citealt{wright2010}) cryogenic and NEOWISE \citep{neowise} post-cryogenic missions. The AllWISE catalog provides the photometry at 3.4 (W1), 4.6 (W2), 12 (W3), and 22 $\mu$m (W4), with the resolutions of 6$\rlap.{''}$1, 6$\rlap.{''}$4, 6$\rlap.{''}$5, and 12$''$, respectively. To convert AllWISE 3.4, 4.6, 12 and 22 $\mu$m magnitudes to fluxes, we use the zero magnitude fluxes of 309.54, 171.787, 31.674, and 8.363 Jy \citep{jarrett2011}, respectively.

\subsection{\textit{Spitzer} MIPS catalog}
\label{sec:mips}

We use complementary MIPS 24 $\mu$m point source catalog provided by the SMOG team, with the angular resolution of $\sim$6$''$. In order to determine the most optimal search radius between the {\it Spitzer}/SMOG IRAC and MIPS catalogs, we performed simple Monte Carlo simulations. In each of 1000 realizations, an artificial SMOG catalog was constructed by shifting all the individual sources in the original catalog in random directions over a distance of 10\arcsec. The artificial catalog is then matched with the actual MIPS data, giving the average number of false matches as a function of the search radius. In order to determine the percentage of incorrect associations in a given sample, the number of false matches is divided by the number of counterparts found using the original catalog. For the {\it Spitzer} MIPS 24\,${\rm \mu m}$ catalog, we adopted a search radius of 1.8$''$, giving a false association rate of 10\%, from which 244 counterparts were found. 

\subsection{\textit{Herschel} EPoS survey}
\label{sec:herschel}

The long-wavelength measurements are taken from the catalog and original images from the Earliest Phases of Star formation (EPoS) {\it Herschel} key program \citep{ragan2012}. The maps of the region were obtained with the Photodetector  Array  Camera and Spectrometer (PACS; \citealt{poglitsch2010}) at 70, 100, and 160 $\mu$m, and the Spectral and Photometric  Imaging REceiver (SPIRE; \citealt{griffin2010}) at 250, 350, and 500 $\mu$m. \iras{} corresponds the EPoS target ISOSS\,J22164$+$6003 and photometry for five point-sources is provided in \cite{ragan2012}. Here, we calculate upper limits for additional sources detected in the IRAS 22147 region using the Level\,3 processed images from the {\it Herschel Science Archive}. We identify point sources as emission peaks and perform photometry by fitting the Gaussian to the source brightness profile using the Curvature Threshold Extractor package (CuTEx; \citealt{molinari2011}). The details of the procedure are described in \citet{molinari2010} and \citet{elia2010,elia2013}. 

\begin{figure*} 
\includegraphics[width=1.0\columnwidth]{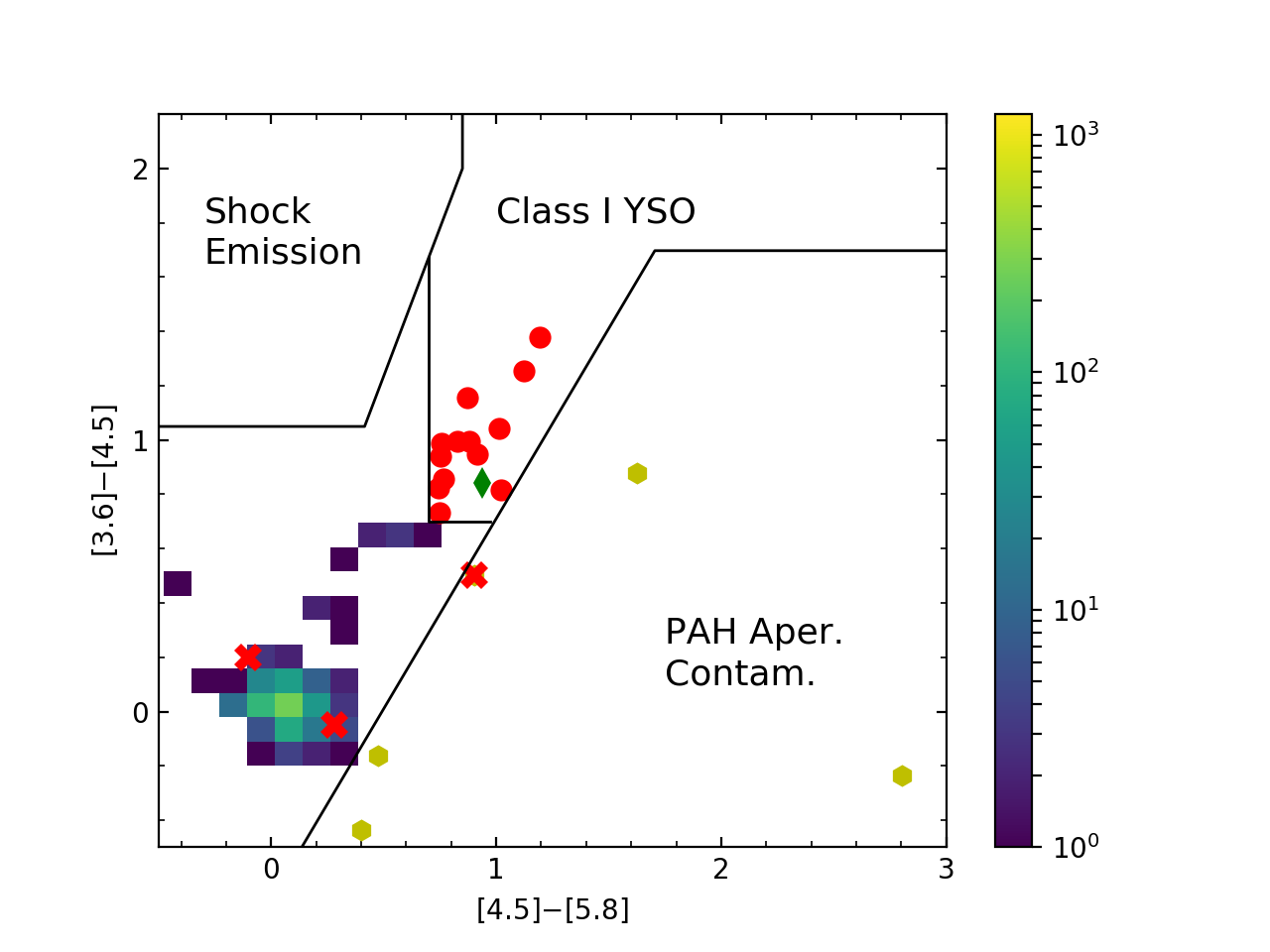}
\includegraphics[width=1.0\columnwidth]{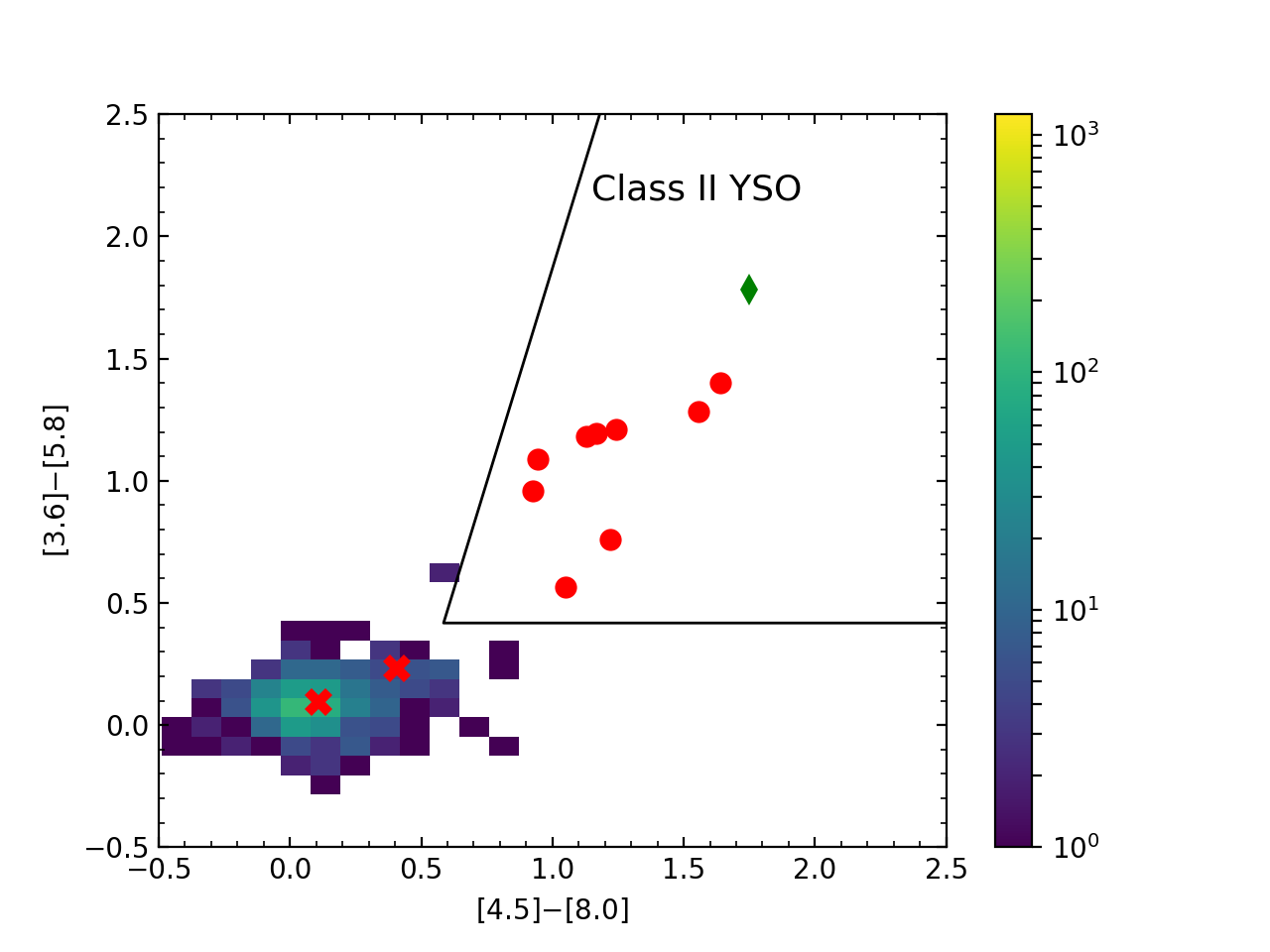} 
\caption{Color-color diagrams using the photometry from \textit{Spitzer}/IRAC. \textbf{Left:} [3.6]-[4.5] vs. [4.5]-[5.8] diagram for the IRAS 22147 region, where all 681 sources with distances and IRAC errors $\sigma<0.2$\,mag are shown. Normal stars are depicted as a Hess diagram (color bar), and excess PAH emission contaminants are shown with light green hexagons and Class I YSOs with red circles. The dark green diamond represents the source initially classified as a Class I YSO, but later re-classified as a heavily reddened Class II source, based on its 24\,${\rm \mu m}$ flux. Red $\times$ symbols indicate three objects identified as YSOs by \citet{Winston2019} but failing to pass our selection criteria (see Section \ref{sec:complet}). \textbf{Right:} [3.6]-[5.8] vs. [4.5]-[8.0] diagram for the IRAS 22147 region, with Class I YSOs and excess PAH emission contaminants from the left panel removed. Symbols are the same as in the left panel.} 
\label{cc1} 
\end{figure*}

\begin{figure} 
\includegraphics[width=1.0\columnwidth]{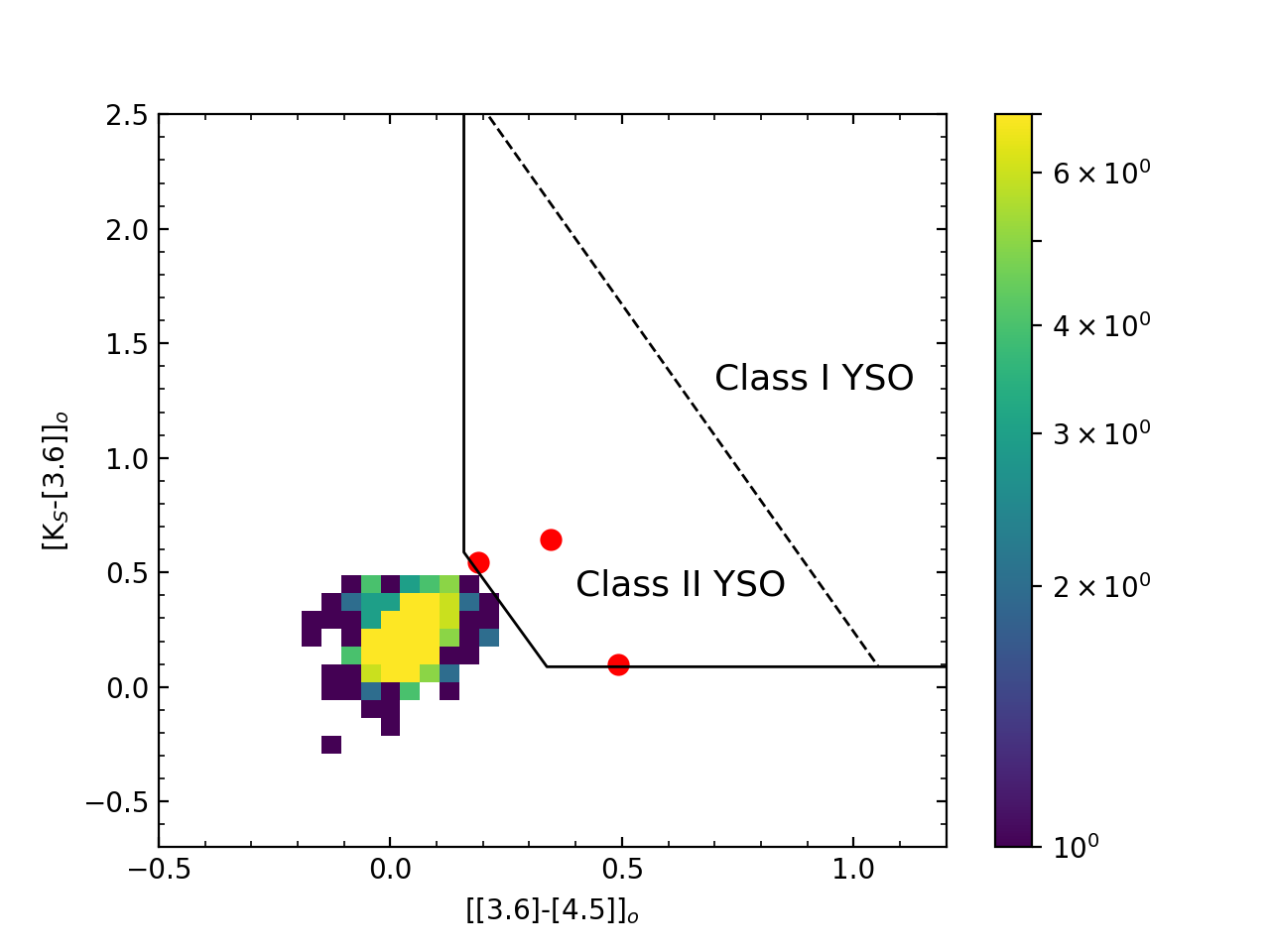} 
\caption{[K$_\mathrm{s}$-[3.6]]$_0$ vs. [[3.6]-[4.5]]$_0$ diagram for the IRAS 22147 region, with  stars and Class II YSOs depicted as Hess diagram and red circles, respectively.} 
\label{cc2}
\end{figure}

\subsection{CO surveys}

We used CO 1-0 data cubes from the Canadian Galactic Plane Survey (CGPS, \citealt{taylor2003}) based on observations from the Five College Radio Astronomy Observatory (FCRAO) CO Survey of the Outer Galaxy \citep{heyer1998}. The CGPS reduction included initial re-processing at DRAO, followed by data resampling and smoothing to the Nyquist resolution limit prior to spatial regridding \citep[see ][for details]{kerton2003}. The spatial resolution is 100.44$\arcsec$ instead of the FCRAO beamsize of 45$\arcsec$. The antenna temperature scale was corrected for forward scattering and spillover losses using a scaling factor of 0.7. 

\subsection{Maser surveys}
\label{sec:maser}
There have been several prior attempts to detect masers towards the IRAS 22147 region, all with negative results. \citet{Palla1991} surveyed the occurrence of the ${\rm H_{2}O}$ 22\,GHz maser emission from a sample of bright IR sources in different star-forming regions. A sample of 260 sources was selected from the IRAS Point Source Catalog, with no detection for \iras{}. \citet{Kalenskij1992} searched for the 7$_{0}$-6$_{1}$ A$^{+}$ CH$_{3}$OH line emission at 44\,GHz with non-detections for \iras{} at the 10\,Jy level. \citet{Wouterloot1993} used 100 m Effelsberg and 32 m Medicina radiotelescopes to search for ${\rm H_{2}O}$ maser emission at 22\,GHz towards 1143 IRAS sources -- again, with no success. Other failed attempts include non-detections of masing lines of SiO \citep{Jiang1996}, NH$_3$ \citep{molinari1996}, OH \citep{Edris2007} and ${\rm CH_{3}OH}$ \citep{Fontani2010}.

In addition, we conducted our own survey at 22\,GHz using Toru\'n 32 m radio telescope. Two series of observations were performed in May 2018 and February 2019, with no detections of 22\,GHz ${\rm H_{2}O}$ maser emission. We used two 4096 channel correlator parts with 8\,MHz bandwidth each, which provided a velocity coverage from $-87$\,km/s to $-33$\,km/s with respect to the local standard of rest. The system temperature was $\sim$75\,K and $\sim$115\,K during first and second part of our survey, which resulted in $3\sigma$ detection limits $\sim$3.5\,Jy and $\sim$5\,Jy, respectively.

\section{Results and analysis} 
\label{sec:res}
In this section, we identify YSO candidates in the IRAS 22147 region using  color-color and color-magnitude diagrams, as well as the machine-learning techniques. We search for YSO candidate clusters and compare the positions of YSO candidates with the spatial extent of the CO 1-0 emission and dust continuum. Finally, we verify the physical properties of the YSO candidates using a modeling of the spectral energy distributions (SEDs).

\subsection{Kinematic distances}
\label{sec:dist}
To find a distance to a given source, we searched for its spatial association with the gas clumps, as traced by the CO 1-0 emission, where we used the data collected as part of CGPS \citep{taylor2003} and FCRAO surveys \citep{heyer1998}.
We searched for the spatially and kinematically coherent structures, namely, the CO 1-0 clumps, using the cloud decomposition techniques, that is, the CLUMPFIND algorithm in the CLOUDPROPS package \citep{Wi94,RL06}. The algorithm considers all the spectral channels along the line of sight of a given source, which are associated with a CO 1-0 clump and assigns a distance to the source based on the clump with the closest centroid \citep{elia2013}, adopting the Milky Way rotation curve of \citet{Br11}. 

Figure \ref{dist} shows a distribution of distances associated with
sources with IRAC photometry with $\sigma<0.2$ identified in the IRAS 22147 region. The histogram shows a bimodal shape with distances clustering around 2.2 kpc and 5.6 kpc.

\begin{figure*} 
\centering 
\includegraphics[scale=0.7]{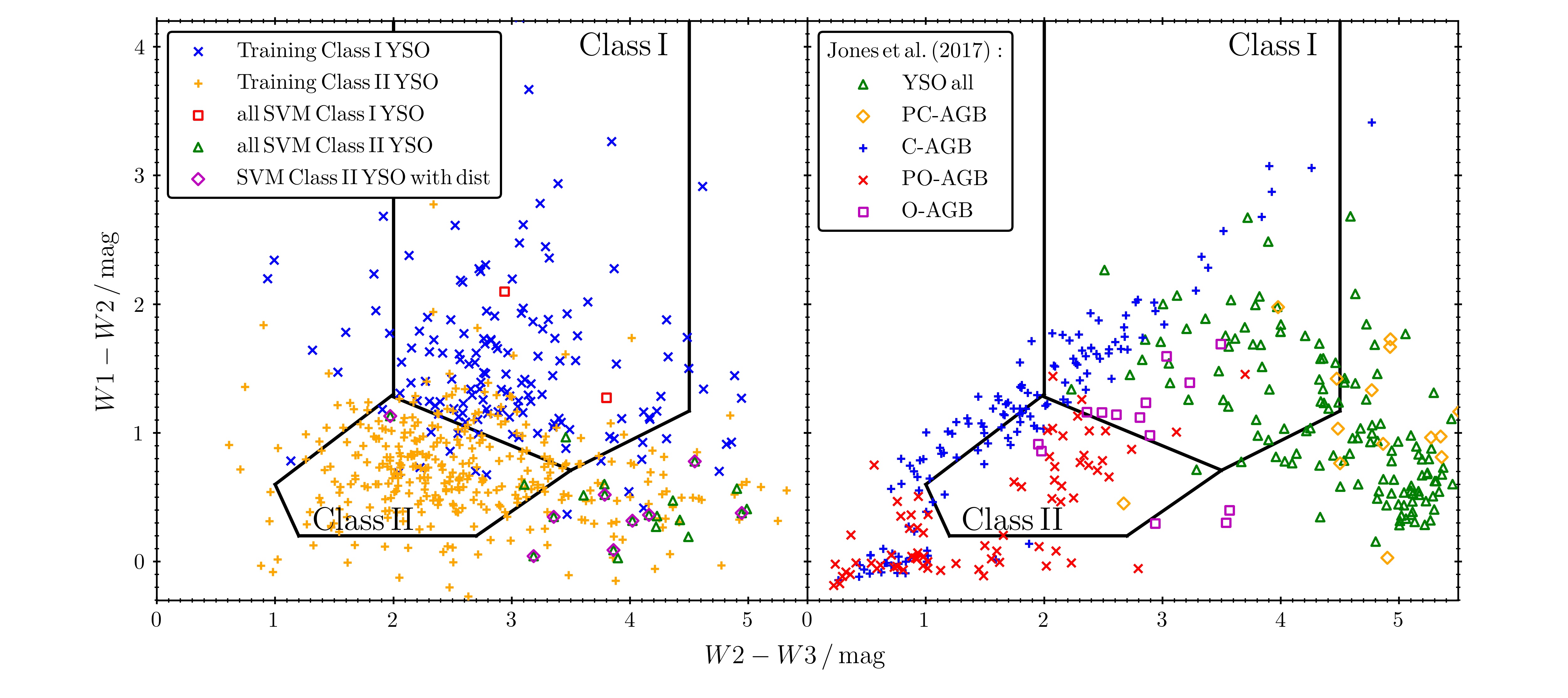}
\caption{\label{w12w23} Color-color diagrams using the AllWISE catalog. \textbf{Left:} $W1-W2$ vs. $W2-W3$ diagram with SVM selected YSO Class I (red squares) and Class II (green triangles) candidates. Pink diamonds mark the position of SVM-selected YSO candidates with measured distances for Class II. Blue $\times$ and orange $+$ symbols show the position of the training YSO I (151 objects) and II (346 objects), respectively, based on the {\it Spitzer} selection in the entire SMOG survey area. Black lines mark the boundaries of  color spaces within which Class I and Class II are expected to lie following the work of \citet{fischer16}. \textbf{Right:} $W1-W2$ vs. $W2-W3$ diagram showing the positions of spectroscopically confirmed YSOs and AGB stars from \citet{jones17}. Green triangles mark all YSOs, yellow diamonds mark the carbon-rich post AGB stars (PC-AGB), blue crosses signs mark the carbon-rich AGB stars (C-AGB), magenta $\times$ symbols mark oxygen-rich post AGB stars (PO-AGB), and pink $\times$ symbols mark the oxygen-rich AGB stars (O-AGB).}
\end{figure*}

The distance of $\sim$5.6 kpc associated with YSO candidates in the central cluster is consistent with the determination by \citet{ragan2012}. The 5\% difference is due to the adopted rotational curve (see Section \ref{sec:sf}). Appendix \ref{sec:gaia} shows a comparison of individual YSO kinematic distances to those obtained using the Gaia DR3 catalog \citep{gaia21}. The distances for two thirds of the sources show good agreement; for the remaining sources, Appendix \ref{sec:gaia} shows the impact of the distance determination on the analysis.

\subsection{Color-color and color-magnitude diagrams}
\label{sec:cc}

To identify YSO candidates using infrared colors, we followed the procedures described in \citet{Gutermuth:2009aa}, where 2548 YSO candidates have been identified inside 36 young, nearby, star-forming clusters, situated at distances between 0.14 and 1.7 kpc.

This approach adopts various color-color (CC) and color-magnitude (CM) cuts and it is, therefore, most reliable when applied to objects with known distances. We note here that \citet{Winston2019} adopted a more liberal version of the \citet{Gutermuth:2009aa} algorithms in order to extend the method to objects without known distances. However, since we do require distances for the later stages of our analysis, namely, the SED fitting and spatial clustering, we strictly follow the more conservative, original approach of \citet{Gutermuth:2009aa}, where we rescaled all the magnitudes to the distance of 1\,kpc.

The starting point of the analysis is the SMOG catalog with 18\,017 sources lying within the IRAS 22147 field, of which 7\,635 ($\sim$42\%) have known distances (see Section\,\ref{sec:dist}). We divide our YSO identification process into three phases, whereby we classify sources based on a number of color-color and color-magnitude cuts. The procedure, described in detail in the appendix of \citet{Gutermuth:2009aa}, can be summarized as follows.

In Phase 1, YSOs were identified based on four \textit{Spitzer}/IRAC bands, where we only considered objects with detections in all four bands with photometric uncertainties $\sigma<0.2$\,mag, resulting in a sample of 681 sources ($\sim$9\%). First, we separated out sources with the excess IR emission that are not YSOs. The active star-forming galaxies were eliminated based on their strong PAH-features with red 5.8 and 8.0\,${\rm \mu m}$ colors. Since these are quite distinct from the YSOs, the contamination from star-forming galaxies in the final sample is expected to be negligible. The broad-line AGNs, with mid-IR colors largely consistent with YSOs \citep{Stern2005}, were removed utilizing the [4.5] vs. [4.5]$-$[8.0] color-magnitude diagram. Finally, the unresolved knots of shock emission, often detected in all IRAC bands, as well as resolved structured PAH-emission producing excess emission in the 5.8 and 8.0\,${\rm \mu m}$ bandpasses, were eliminated. The YSOs were then selected from the cleaned sample, where we found 14 Class I and 9 Class II sources (Figure\,\ref{cc1}).

Phase 2 was applied to sources lacking 5.8 or 8.0\,${\rm \mu m}$ detections, but having high quality ($\sigma<0.1$\,mag) 2MASS data in $H$ and $K_s$ ($J$ can also be used when present). Out of 280 sources passing those criteria, only 3 are located within the Class II region of the diagram (Figure\,\ref{cc2}). 

Finally, in Phase 3, we utilized the MIPS 24\,${\rm \mu m}$ data in order to re-examine the original catalog. In this step we find one of our Class I YSOs from Phase 1 to lack adequate 24\,${\rm \mu m}$ excess, which reclassifies it as a heavily reddened Class II YSO (green diamond in Figure\,\ref{cc1}). We therefore ended up with 26 YSO candidates, namely, 13 of Class I and 13 of Class II. 

In order to maximize the YSO selection process by including the available AllWISE data, we utilized our YSO candidate sample as a training set for a machine learning algorithm, as described in the following section.

\subsection{Machine learning technique}
\label{sec:ml}

Machine learning algorithms are a branch of artificial intelligence whereby the algorithm is designed to learn from data, identify patterns or structures within the data, and then make predictions with minimal human involvement. Here, we used a combination of the machine learning anomaly detection and supervised algorithms to identify candidate YSOs in the IRAS 22147 region. We followed procedures developed specifically for the analysis of the AllWISE catalog, which led to the identification of, for example, heavily reddened AGN and quasar candidates that were originally missing from source catalogs created using standard cuts on CC diagrams \citep{solarz17}. The process of selection is applied to the entire SMOG region, but the output catalog discussed here includes only the IRAS 22147 region.

As a first step, we distinguished sources with YSO signatures from the rest of the Galactic sources. For that purpose, we used the anomaly detection algorithm called the one-class support vector machines (OCSVM, \citealt{ocsvm}) with the R interface for libsvm \citep{man1,man2}. In particular, OCSVM is designed to select outliers in the data based on the input data, where all known objects have a single label: "known." As the training objects, we used all the YSOs selected through the Spitzer's color-color diagram method (Section\,\ref{sec:cc}) that have a counterpart in the AllWISE data (151 Class I and 346 Class II objects).

The training process of the OCSVM starts with the creation of a feature vector associated with each training point, composed of $N$ quantities that describe the discriminating properties of a given object. In case of the AllWISE data, we use 4 AllWISE fluxes and three color combinations ($W1-W2$, $W2-W3,$ and $W3-W4$), so, in total, a 7D input parameter space. Subsequently, these numbers would act as coordinates for each training object (vector) in 7D feature space. Using the pre-defined training sets of sources, the algorithms used the kernel functions to map the input parameter space into a higher dimensional feature space  \citep{shawetaylor04}. The algorithm will search for a hyperplane, which separates the training points and creates a "normality" model. Once this process is complete, the new data, which does not match the "known" object patterns, are flagged as outliers.

In the IRAS 22147 region, we focus on sources which display the behavior similar to known YSOs. Therefore, we use the OCSVM to perform a "reversed" anomaly detection, that is, we train the algorithm on the known YSOs and we reject all those sources which do not fit the model created based on the "known" YSOs. Using the sample of general YSO candidates selected by the OCSVM, we proceeded to classify the sources as either Class I or II. For that purpose, we used the classic support vector machine algorithm, which is designed to recognize two (or more) types of objects based on the training examples provided by the supervisor (SVM, \citealt{vapnik95}). 

The process is similar to the OCSVM, except that the algorithm searches for the most optimal hyperplane separating the examples of the two categories, instead of searching for the most optimal enclosure for these sources. The hyperplane separates the examples from two categories with the largest possible margin. We used the same training sample as for the anomaly detection, but divided into separate Class I and II groups. 

The number of Class I and II sources detected in the IRAS 22147 region using the machine learning techniques is 2 and 21, respectively. Out of these sources, only 9 Class II sources are assigned kinematic distances. 
The left panel of Figure \ref{w12w23} shows the YSO candidates in the $W1-W2$ vs. $W2-W3$ diagram. All Class I and II candidates identified with the machine-learning algorithms
 are located in the parts of the diagram, where the training sample obtained using the color-color and color-magnitude diagrams is also located (Section \ref{sec:cc}).
 The right panel of Figure \ref{w12w23} shows the WISE colors of different types of dusty objects
  from Large Magellanic Cloud (LMC), classified by \cite{jones17} based on the \textit{Spitzer}/IRS spectroscopic observations. The \cite{jones17} catalog includes evolved objects: carbon and oxygen
   rich asymptotic giant branch (AGB) and post-AGB stars (C-AGB, O-AGB, PC-AGB and PO-AGB, respectively), as well as different types of YSOs. We can see that the majority of the spectroscopically
    observed YSOs in LMC lie outside of the \citet{fischer16} YSO selection lines.
    
The SVM method selects YSO candidates, which lie in the vicinity of the border proposed for Class II YSOs, with the $W2-W3$ color reddened by up to 2 mag \citep{fischer16}.
 A similar area of the CCD is occupied by confirmed YSOs from Taurus \citep{Koenig2014} and star-forming regions in the LMC \citep{jones17}; however, it might be to some extent contaminated
  by star-forming galaxies. Thus, for the sake of catalog reliability, the region with numerous galaxies is not recommended
  for the YSO classification based solely on the IR photometry \citep{Koenig2014,fischer16}. Here, we find that all the SVM selected candidate YSOs are associated with CO emission (see Section
   \ref{sec:spatial}), which implicates that they are likely to be well-identified by the algorithm. However, a contamination from the PC-AGB stars could be still present.

\subsection{Comparison to Winston et al. (2019) and completeness}
\label{sec:complet}

Adopting color-color and color-magnitude cuts, as well as machine learning techniques, we end up with a sample of 49 YSO candidates, 15 Class I and 34 Class II (see Appendix B for their photometry). \citet{Winston2019}, within the IRAS 22147 region, identified 49 YSOs, 16 Class I, 29 Class II, and 4 Class III. In this section, we discuss some of the differences between both samples and completeness of the final sample.

There are 24 YSOs, which were identified in both samples. Thirteen Class I and 11 Class II objects have been identified in this work, while \citet{Winston2019} finds 11 Class I and 13 Class II sources (see Appendix C for a detailed comparison). The small discrepancies in the classification of the identified YSOs can be attributed to the subtle differences in the color-color and color-magnitude cuts applied in both works. 

\begin{table} 
\caption{Clusters of YSOs in the IRAS 22147 region\label{clust}}          
\centering                        
\begin{tabular}{ccccccc}
\hline \hline                 
$d$ & $l$ & $b$ & $N_{\rm I}$ & $N_{\rm II}$ & $N_{\rm II}/N_{\rm I}$ \\
(kpc)  & $(^{\circ})$ & $(^{\circ})$ &  &  &  \\
\hline
2.207 &104.726&2.605&1&4&4.00\\
2.215 &104.762&2.717&2&2&1.00\\
2.223 &104.913&2.771&2&0&--\\
5.628 &104.709&2.797&3&8&2.67\\
5.665 &104.816&2.699&3&1&3.00\\
\hline 
\end{tabular}
\begin{flushleft}
Coordinates ($l$, $b$) correspond to the location of the center of the clusters. $N_{\rm I}$ and $N_{\rm II}$ are numbers of Class I and Class II YSO candidates in the clusters, respectively.
\end{flushleft}
\end{table}

\begin{figure}
\centering
\includegraphics[scale=0.67]{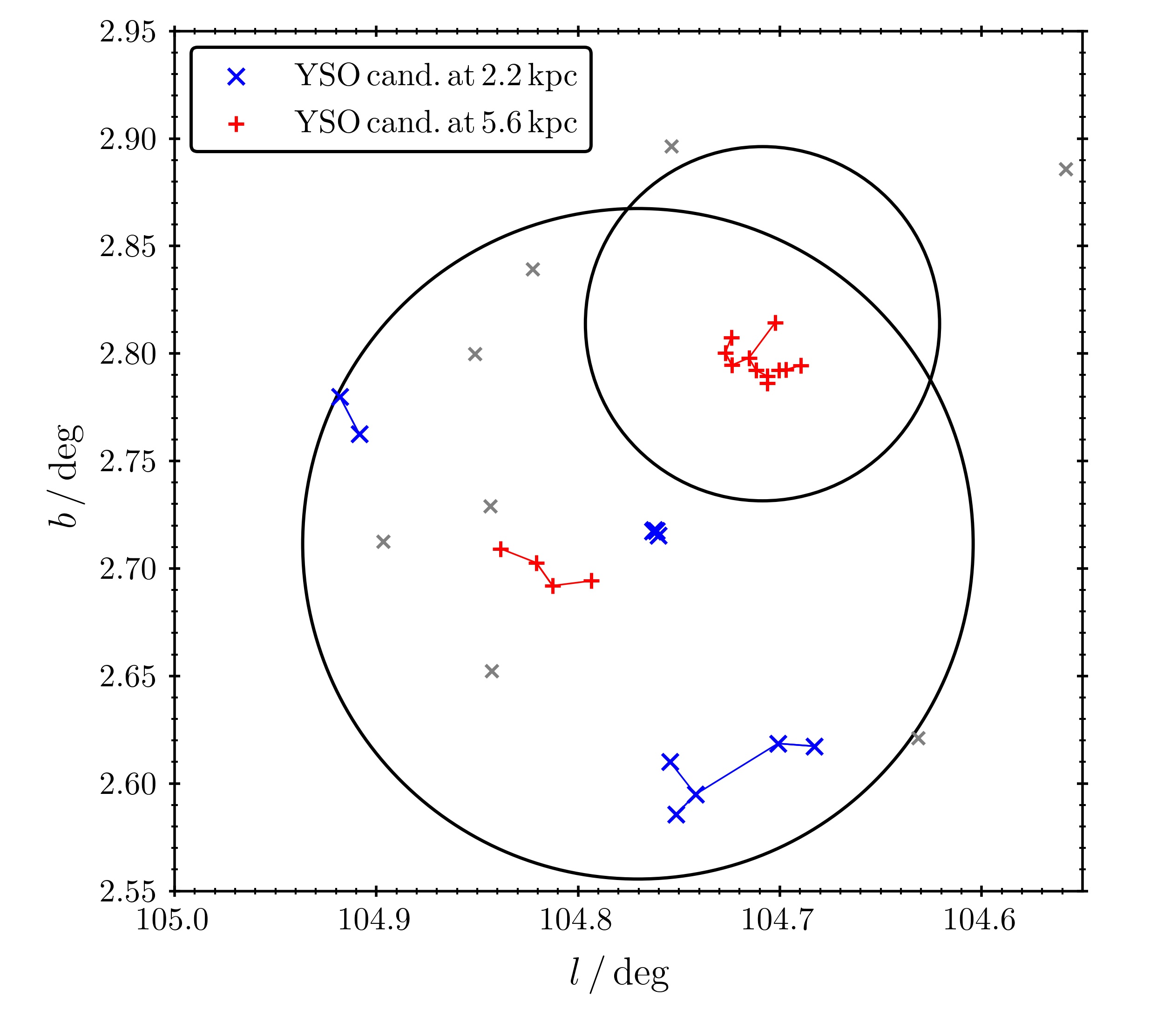}
\caption{Minimum spanning tree for YSO candidates in the IRAS 22147 region. Blue and red $\times$ symbols show objects belonging to clusters located at $d\sim 2.2$ and $d\sim 5.6$~kpc, respectively; the isolated objects are shown as gray $\times$ symbols. Blue and red lines correspond to branches connecting objects from the same clusters. Black circles correspond to the two clusters identified in the region by \citet{Winston2019}.}
\label{mst}
\end{figure}

\label{sec:assocs}
\begin{figure*}
\begin{center} 
\includegraphics[scale=0.67, trim=0cm 0cm 0cm 0cm]{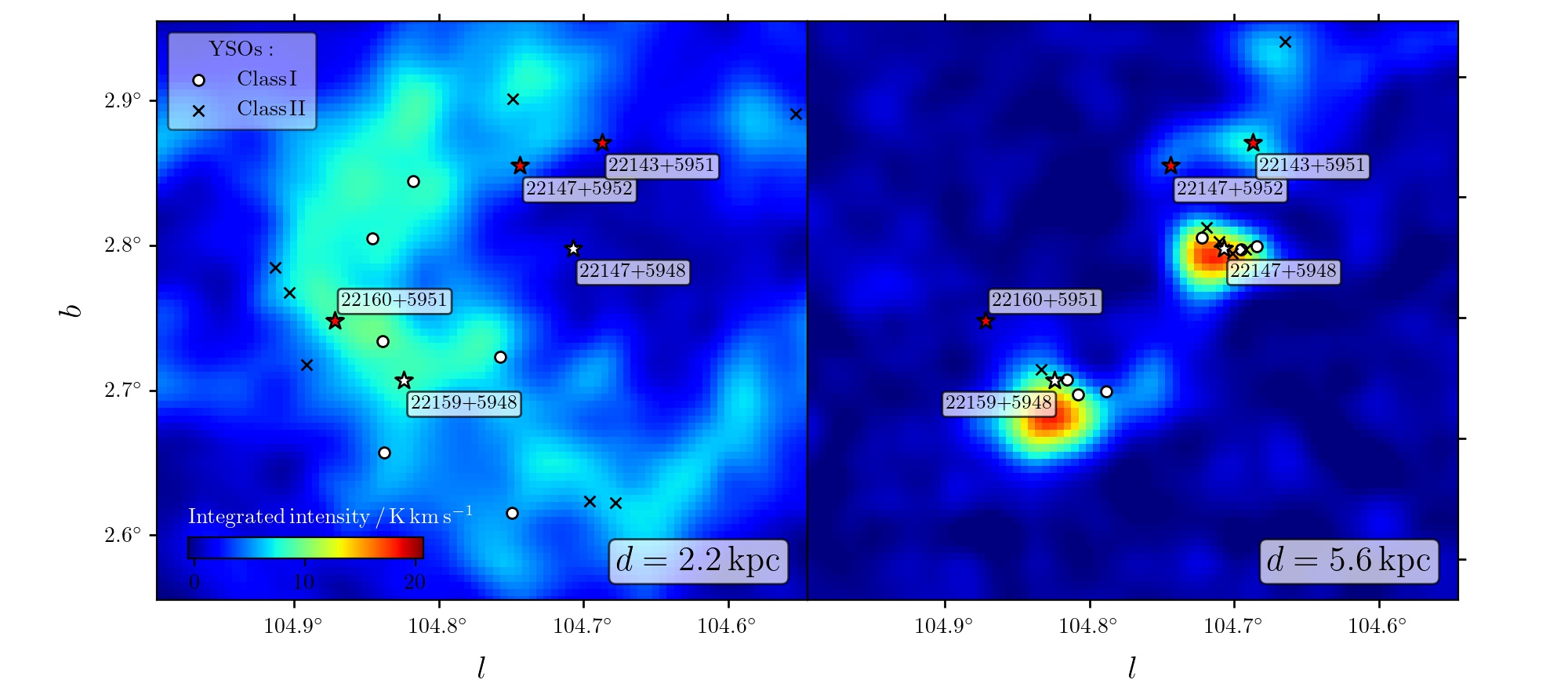}
\end{center} 
\caption{Integrated intensity CO 1-0 maps from FCRAO using velocity ranges from -25.4 km s$^{-1}$ to -23.7 km s$^{-1}$  (left) and -60.8 km s$^{-1}$ to -58.4 km s$^{-1}$ (right). The positions of YSOs, as classified via CCD/ML methods (Sections\,\ref{sec:cc} and \ref{sec:ml}), are shown in circles (Class I) and $\times$ symbols (Class II). Positions of \iras{} and IRAS 22159+5948 are shown in white stars, and those of other IRAS sources in the region as red stars.}
\label{fig:co} 
\end{figure*}

\begin{figure*} 
\begin{center} 
    \includegraphics[scale=0.67, trim=0cm 1cm 0cm 0cm]{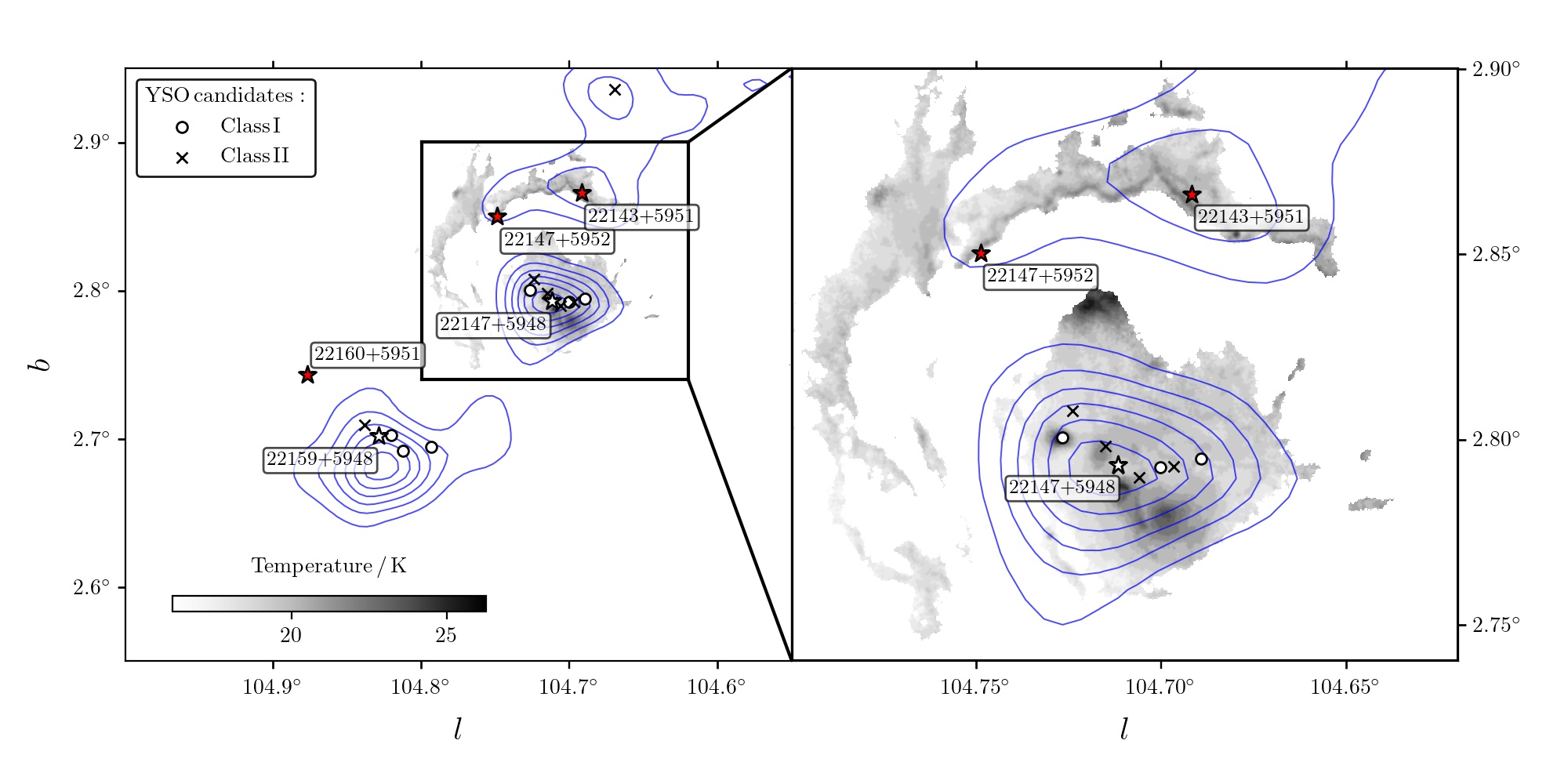}
\end{center} 
\caption{Dust temperature map (grayscale) and CO 1-0 map (blue contours) for the gas components at the distance of 5.6 kpc. The dust temperature distribution is limited to the central region, covered by the {\it Herschel} EPoS key program (Section\,\ref{sec:herschel} and Figure\,\ref{fig:st}) and, therefore, does not cover the south-east feature. The CO contours start at 2.7 K km s$^{-1}$
(5 $\sigma$) and are drawn in steps of 2.7 K km s$^{-1}$; the range of CO integrated intensities is shown on Fig. 7. The positions of YSOs, as classified via CCD/ML methods (Sections\,\ref{sec:cc} and \ref{sec:ml}), are shown in circles (Class I) and $\times$ symbols (Class II). }
\label{fig:temp} 
\end{figure*}

Twenty five sources identified in \citet{Winston2019} do not appear in our sample:\ 9 of them have errors in {\it Spitzer}/IRAC bands larger than 0.2 mag, while 13 were not assigned distances, which immediately removes them from our sample. One was classified as a resolved structured PAH-emission source, while the remaining 2, again, were not identified in this work due to differences in cuts used. We show them as red crosses in Figure\,\ref{cc1}.

Finally, 25 sources have been classified as YSOs in this work but do not appear in \citet{Winston2019}. Eighteen of them have been classified using the AllWISE data. While \citet{Winston2019} uses color-color cuts in 4 WISE bands from \citet{Koenig2014} and \citet{fischer16} (black lines in Figure\,\ref{w12w23}), we adopted the machine learning algorithms (Section\,\ref{sec:ml}) based on the training sample of candidate YSOs selected using IRAC color-color and color-magnitude diagrams. In the left panel of Figure\,\ref{w12w23} we can see that out of 21 sources classified as YSO Class II using ML techniques (green triangles), 18 fall outside the region of the color-color diagram used by \citet{Winston2019}. The remaining 7 YSOs that do not appear in \citet{Winston2019} have been identified in this work using the IRAC CC/CM diagrams. As above, this is caused by the \citet{Winston2019} adjusting the original \citet{Gutermuth:2009aa} cuts to objects at larger distances, which resulted in subtle cut differences.

Apart from the differences in the identification methods, the number of YSO detections depends on the sensitivity of the instruments and their capability to detect point-sources at a given distance. In the case of Class 0/I protostars, an empirical relation between the sensitivity in luminosity and the distance of the object has been proposed by \cite{dun06,dun08}, and subsequently confirmed for objects in the outer Galaxy by \cite{ragan2012}. Accordingly, the limiting luminosities of 1.0 and 6.4 L$_{\odot}$ are expected for Class 0/I sources at $\sim$2.2 kpc and $\sim$5.6 kpc, respectively, in agreement with bolometric luminosities (Appendix \ref{sec:lbol}).

Additional limitation on the detection of the most embedded, namely, the least evolved, YSOs stem from the adopted CC cuts, which require that the target is detected in all four \textit{Spitzer}/IRAC bands. This could be the case for Class 0 YSOs, whose SEDs are dominated by the cold, dense envelopes with the peak around 100 $\mu$m, and very little or no near-IR emission is observed \citep{andre93,evans2009,karska2018}. The \textit{Herschel}/PACS spectral maps allowed the identification of five deeply-embedded YSOs in the IRAS 22147 region with $L_\mathrm{bol}$ from 24 to 260 L$_{\odot}$ \citep{ragan2012}. Two sources, G104.6895+02.7945 and G104.7270+2.8003, are identified as Class I YSOs using the CC diagram method presented here; however, the three remaining sources -- including \iras{} -- lack the near-IR photometry in the \textit{Spitzer} point source catalog. Similarly, two IRAS sources associated with an arc of emission to the north of \iras{} also lack the \textit{Spitzer}/IRAC emission and are not classified here.

\begin{table*} 
\caption{Best-fit models of the spectral energy distribution using the \citet{robitaille2017} classification \label{t:modelstats}}
\centering                        
\begin{tabular}{cclc}
\hline \hline                 
Model Set & \# of sources & Components & Group \\ 
\hline  
sp$--$s$-$i & 12 & star $+$ passive disk; $R_{\rm inner}$ = $R_{\rm sub}$ & $d$\\
sp$--$h$-$i & 4 & star $+$ passive disk; variable $R_{\rm inner}$ & $d$ \\
sp$--$smi & 5 & star $+$ passive disk $+$ medium; $R_{\rm inner}$ = $R_{\rm sub}$ & $d$ \\
sp$--$hmi & 2  & star $+$ passive disk $+$ medium; variable $R_{\rm inner}$ & $d$ \\
s$-$ubsmi & 1 & star $+$ Ulrich envelope $+$ cavity $+$ medium; $R_{\rm inner}$ =$R_{\rm sub}$ & $e$\\
spu$-$hmi & 1 & star $+$ passive disk $+$ Ulrich envelope $+$ medium; variable $R_{\rm inner}$ & $d+e$\\
spubsmi & 1 & star $+$ passive disk $+$ Ulrich envelope $+$ cavity $+$ medium; $R_{\rm inner}$ = $R_{\rm sub}$ & $d+e$\\
\hline 
\end{tabular} 
\begin{flushleft} The model set names are from \citet{robitaille2017}. Seven characters in the model set names indicate which component is present; they are (in order): s: star; p: passive disk, p or u: power-law or Ulrich envelope; b: bipolar cavities; h: inner hole; m: ambient medium; and i: interstellar dust. A dash ($-$) is used when a component is absent.\\
$R_{\rm inner}$ is the inner radius for the disk, envelope, and the ambient medium - when one or more of these components are present. $R_{\rm sub}$ is the dust sublimation radius.\\
\end{flushleft}
\end{table*}
\begin{figure*} 
\begin{center} 
\includegraphics[angle=0,height=6cm]{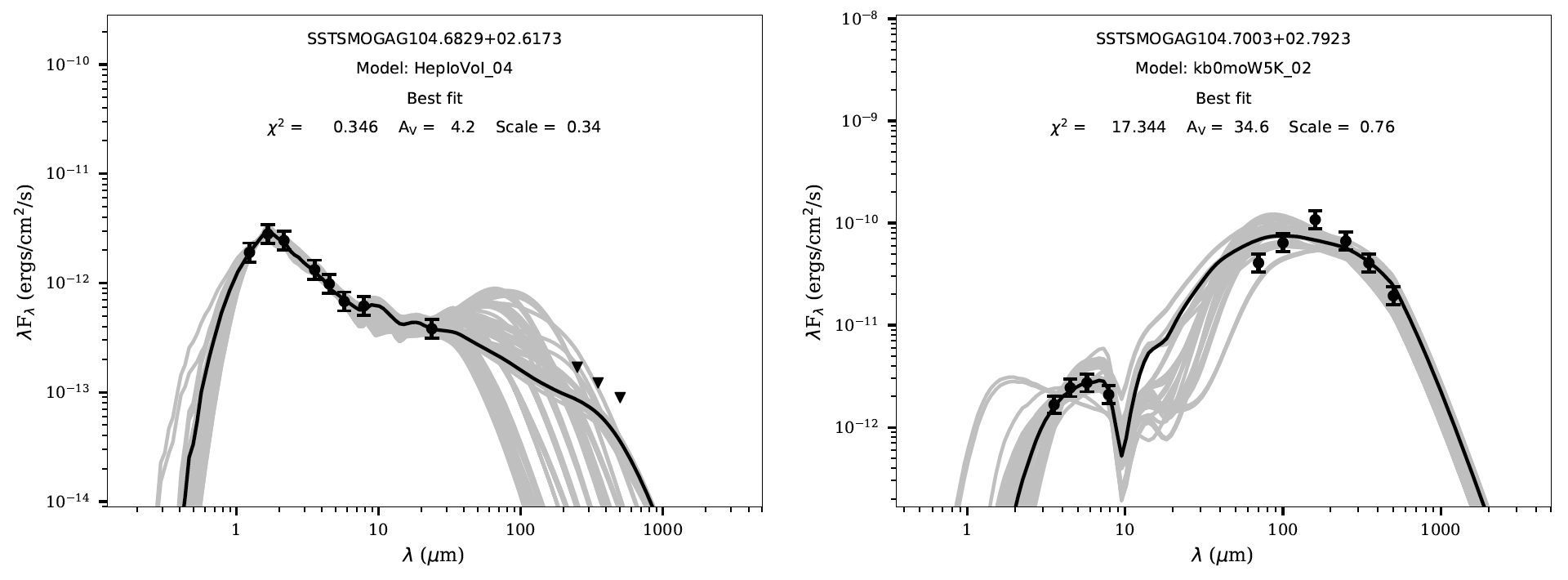}
\end{center} 
\caption{Example SEDs of candidate YSOs with well-fit \cite{robitaille2017} YSO models. The best fit model is indicated with the black solid line; gray lines show the YSO models with $\chi^{2}$ between $\chi^{2}_\mathrm{best}$ and $\chi^{2}_\mathrm{best}$+$F$, where $F$ is a threshold parameter which we set to 3 \citep{sewilo2019}. Filled circles and triangles are valid flux values and flux upper limits, respectively. The values of a reduced $\chi^{2}$ and interstellar visual extinction for the best-fit model are indicated in the plots. Appendix \ref{sec:seds2} shows the SEDs for the remaining YSOs in the IRAS 22147 region.}
\label{seds} 
\end{figure*}

\subsection{Spatial clustering of selected YSOs}
\label{sec:clust}

The spatial distribution of YSO candidates provides support for their physical association and likely a shared origin. For the case of the Outer Galaxy sources, it is yet another means of avoiding the cloud confusion and confirming the YSO identification.

To identify clusters in the IRAS 22147 region, we used the agglomerative hierarchical clustering algorithm \citep{everitt11}, which belongs to the unsupervised family of machine learning. Similar to all clustering algorithms, it serves to classify a set of objects into subgroups that display similar characteristics. The definition of the similarity measure among objects depends on the metrics definition.

As a first step, the algorithm assigns each data point in its own cluster. Then, it identifies the closest two clusters and combines them into one. This process is re-iterated until all data points reside within a single cluster. After the YSOs have been combined into a single cluster, the user chooses a critical length and a minimum membership number. The process can be represented by a dendrogram structure showing a hierarchical relationship between objects. The advantage of the adopted clustering algorithm is the fact
that the user does not have to define the number of classes a priori, but can choose a fitting number once dendrograms are calculated. Following \cite{Gutermuth:2009aa}, we employed single-linkage agglomerative clustering  where links longer than the critical length are cut, and the remaining clusters with at least a minimum number of members are preserved.  

We calculated the projected distances in a standard way using source coordinates, but separately for sources at distances of $\sim$2.2 and 5.6 kpc. The exact distance regimes are chosen based on the distance histogram (Fig. 2), with sources at the distance from 1.2 to 2.4 kpc referred to as 2.2 kpc, and those with distances above 4 kpc referred to as 5.6 kpc (Fig. ~\ref{mst}). Due to a visual inspection of dendrograms, a critical length of 170 and 100 arcseconds was set for objects at roughtly 2.2. and 5.6 kpc, respectively. 

 Figure ~\ref{mst} shows the minimum spanning tree for clusters associated with \iras{}.
Five clusters containing between 2 and 11 members, and a total of 26 YSO candidates, are identified (see also Table 1). Eight sources are not associated with any of the clusters. We note that there are three clusters at $\sim2.2$ kpc and one cluster at $\sim5.6$ kpc covering the area of a single cluster identified in \cite{Winston2019}, using the Density-Based Spatial Clustering and Application with Noise \citep[DBSCAN, ][]{ester1996}, see Section \ref{sec:algo}. In their analysis, no distance information were available, causing the line-of-sight confusion (see Fig. \ref{mst}).

\citet{fischer16} note that clusters with fewer than five members should be discounted as unrealistic chains and may be produced by the single-linkage method. Accordingly, we would have two fully reliable clusters corresponding to two distance regimes with four and with eight members. Yet, there are clearly two separate clusters at 5.6 kpc in the considered region (Figure 1), corresponding to \iras{} in N-W and IRAS 22159+5948 in S-E. These two clusters are likely physically linked, but nevertheless, we chose to keep them separate in our identification. 

\subsection{Spatial extent of CO gas and dust} 
\label{sec:spatial}

Young stellar objects at early evolutionary stages are associated with gas and dust from their nascent molecular cloud cores. Extended CO 1-0 emission pin-points gas in the envelopes and outflows from Class 0/I protostars. Dust emission traces star forming filaments, cores and envelopes. Appendix E shows images of the IRAS region in individual bands from near-IR to radio wavelengths.

Figure \ref{fig:co} shows the spatial distribution of CO 1-0 emission, separately for the two line components which correspond to distances of $\sim$2.2 and 5.6 kpc (see Section\,\ref{sec:dist}). The CO gas at $\sim$2.2 kpc forms an extended arc to the east and south from \iras{}. The component at $\sim$5.6 kpc shows two gas concentrations which corresponds to the positions of \iras{} and IRAS 22159+5948. Additionally, a weak arc of emission is detected to the north of \iras{} and a bridge in between the two IRAS sources. 

The positions of YSO candidates identified in Section \ref{sec:cc} and \ref{sec:ml} closely follow the distribution of CO 1-0, supporting their YSO classification; YSOs at $\sim2.2$ kpc are consistent with a few small clusters which are rather loosely distributed (see Section \ref{sec:clust}). The objects at $\sim$5.6 kpc form well-defined clusters, consisting of Class 0/I candidates, as opposed to the Class II and III populations found at 2.2 kpc.

The spatial resolution of the CGPS CO 1-0 cube of the order of 100$\arcsec$ does not allow us to isolate emission from individual objects within the cluster associated with \iras{}. Yet, in locations corresponding to clusters of YSOs, emission in the CO 1-0 line wings has been detected, which is a signature of outflows and the young evolutionary status of the driving sources (Class 0/I). Higher spectral and angular resolution observations are necessary to characterize outflows and envelopes from individual YSOs. 
\begin{table*} 
\caption{Physical parameters for a subset of YSO Candidates in the IRAS 22147 region with at least five photometry points, including one at a wavelength longer than 12\,${\rm \mu m}$ \label{t:physpar}}          
\centering                        
\begin{tabular}{rcccccccccc}
\hline \hline                 
Nr & IRAC Designation & Model & Class$^b$ & Class$^c$ & Selection & $R_{\rm \star}$ & $T_{\rm \star}$ & $L_{\rm \star}^{e}$ & $M_{\rm \star}^{f}$ & Age$^f$ \\
 & `SSTSMOGA'        & Set$^a$ & init    & SED      & method$^d$ & ($R_{\odot}$)  & (K)   & ($L_{\odot}$)   & ($M_{\odot}$)       & (Myr)   \\
\hline
\multicolumn{11}{c}{YSO candidates at $\sim$5.6 kpc} \\
\hline
1 & G104.6697+02.9357 & sp--smi & II & II+III & ML &  $2.8$ & $11340$ & $120$ & 3.0 & 2.3 \\ 
2 & G104.6895+02.7945 & sp--smi & I & II+III & CC &  $54.5$ & $6760$ & $5600$ & -- & -- \\ 
3 & G104.6968+02.7925 & s-ubsmi & II & 0 & ML &  $4.2$ & $10690$ & $210$ & -- & -- \\ 
4 & G104.7003+02.7923 & spubsmi & I & I & CC &  $16.4$ & $5070$ & $160$ & -- & -- \\ 
5 & G104.7062+02.7895$^g$ & sp--h-i & II & II+III & ML &  $32.9$ & $6500$ & $1750$ & -- & -- \\ 
6 & G104.7118+02.7921 & sp--h-i & II & II+III & ML &  $58.9$ & $4050$ & $840$ & -- & -- \\ 
7 & G104.7153+02.7979 & sp--h-i & II & II+III & ML &  $5.4$ & $11280$ & $430$ & 4.4 & 0.8 \\ 
8 & G104.7241+02.8075 & sp--h-i & II & II+III & ML &  $11.2$ & $5630$ & $110$ & -- & -- \\ 
9 & G104.7270+02.8003 & sp--smi & I & II+III & CC &  $11.5$ & $14370$ & $5050$ & -- & -- \\ 
10 & G104.7934+02.6942 & sp--s-i & I & II+III & CC &  $3.6$ & $10150$ & $120$ & 3.0 & 2.2 \\ 
11 & G104.8126+02.6920 & sp--s-i & I & II+III & CC &  $6.4$ & $11760$ & $700$ & -- & -- \\ 
12 & G104.8206+02.7026 & sp--smi & I & II+III & CC &  $1.5$ & $6980$ & $5$ & -- & -- \\ 
13 & G104.8383+02.7091 & sp--s-i & II & II+III & CC &  $13.8$ & $4790$ & $90$ & -- & -- \\ 
\hline  
\multicolumn{11}{c}{YSO candidates at $\sim$2.2 kpc} \\
\hline
1 & G104.5582+02.8857 & sp--s-i & II & II+III & CC &  $2.8$ & $5560$ & $7$ & 2.1 & 2.4 \\ 
2 & G104.6829+02.6173 & sp--s-i & II & II+III & CC &  $2.0$ & $4220$ & $1$ & 0.7 & 1.0 \\ 
3 & G104.7009+02.6186 & sp--s-i & II & II+III & CC &  $1.1$ & $5770$ & $1$ & -- & -- \\ 
4 & G104.7538+02.8964$^g$ & sp--hmi & II & II+III & ML &  $1.4$ & $6490$ & $3$ & -- & -- \\ 
5 & G104.7544+02.6101 & sp--s-i & I & II+III & CC &  $6.7$ & $4160$ & $12$ & -- & -- \\ 
6 & G104.7622+02.7179 & spu-hmi & I & I & CC &  $4.1$ & $4400$ & $6$ & 0.8 & 0.2 \\ 
7 & G104.8225+02.8393 & sp--s-i & I & II+III & CC &  $1.5$ & $5150$ & $2$ & 1.4 & 4.1 \\ 
8 & G104.8429+02.6522$^g$ & sp--s-i & I & II+III & CC &  $0.6$ & $3840$ & $0.1$ & -- & -- \\ 
9 & G104.8436+02.7289 & sp--s-i & I & II+III & CC &  $1.4$ & $4450$ & $0.6$ & -- & -- \\ 
10 & G104.8510+02.7998 & sp--s-i & I & II+III & CC &  $2.9$ & $5580$ & $7$ & 2.1 & 2.3 \\ 
11 & G104.8965+02.7124 & sp--smi & II & II+III & ML &  $1.9$ & $5650$ & $3$ & -- & -- \\ 
12 & G104.9084+02.7625$^g$ & sp--s-i & II & II+III & CC &  $2.5$ & $8270$ & $27$ & -- & -- \\ 
13 & G104.9180+02.7798 & sp--hmi & II & II+III & ML &  $0.9$ & $4800$ & $0.4$ & -- & -- \\ 
\hline 
\end{tabular}
\begin{flushleft}
$^a$See Table~\ref{t:modelstats} footnotes for the meaning of characters in the model set names.\\
$^b$YSO class as initially determined using color-color diagrams and machine learning algorithms (Sections \ref{sec:cc} and \ref{sec:ml}, respectively).\\
$^c$YSO class as determined from the SED fitting (see Section\,\ref{sec:seds}).\\
$^d$Initial selection methods used in this work are the color-color/color-magnitude diagrams (CC; Section\,\ref{sec:cc}) and machine learning algorithms (ML; Section\,\ref{sec:ml}).\\
$^e$ Stellar luminosities are calculated using the formula $L_{\star}=4\pi R_{\star}^2\sigma_{SB}T_{\star}^4$, where $\sigma_{SB}$ is the Stefan-Boltzmann constant.\\
$^f$We provide stellar masses ($M_{\rm \star}$) and ages ($Age$) only for sources with the most reliable estimates of these parameters (see Section\,\ref{sec:seds}).\\
$^g$ A more reliable distance for this source is provided by Gaia; see Appendix \ref{sec:gaia} for the SED model adopting the Gaia DR3 distance from \cite{gaia21}. \\
\end{flushleft}
\end{table*}

Figure \ref{fig:temp} shows the spatial distribution of the dust continuum emission in the far-infrared, its temperature, and a comparison with the CO 1-0 emission (see also Figure \ref{fig:co}). Observations available from the EPoS survey cover only the west part of the region containing the cluster associated with \iras{}. 

Emission from the dust is co-spatial with CO gas detected at $\sim$5.6 kpc, strongly suggesting that the most prominent structures seen in Figure 1 all belong to \iras{} and not the star-forming region at $\sim$2.2 kpc. Consequently, the dust also follows the positions of YSO candidates detected in the cluster, as well as the northern arc of material.

The dust temperatures were calculated using the photometry from {\it Herschel} at 70 and 160 $\mu$m, assuming a power-law dust emissivity (Equation \ref{eq1}), where $B_\nu$ is the Planck function and $\beta$ is the emissivity index: 
\begin{equation}
\label{eq1}
I_\nu\propto \lambda^{-\beta}B_\nu(T_{\rm dust}).
\end{equation}

We adopted a constant value of $\beta$ equal to 1.7, which is typical for envelopes of Class 0 and I YSOs \citep{Kr12,Go12}. The dust temperatures range between 15 and 30 K, in agreement with typical temperatures in sites of low-mass star formation \citep{Kr12}. Temperature peaks of $\sim25$ K typically omit the positions of YSO candidates with one exception, corresponding to evolved Class II and III candidates at the far-east of the cluster. The temperature increase is most prominent at the northern edge of the cluster, which is devoid of YSO candidates found in Sections \ref{sec:cc} and \ref{sec:ml}. This region corresponds to the bright MIPS 24 $\mu$m emission in Figure 1, which contains several sources that were not identified as YSOs.

To summarize, the spatial extent of CO gas and dust provides important implications for the region and help distinguish star forming regions along the same line-of-sight. The distance calculated from CO emission verifies the clustering in \iras{} and favors the interpretation of two small clusters at 5.6 kpc (Section \ref{sec:clust}) instead of the cluster identification proposed by \citet{Winston2019}. The CO 1-0 line wings detected toward the IRAS 22147 region confirm the formation of deeply-embedded protostars in the region. The gas and dust distribution together explain the composite mosaics images and imply the physical connection of the observed structures. 

\subsection{Spectral energy distributions}
\label{sec:seds}

    An additional test of the YSO identification and classification is available via the modeling of their broad-band SEDs. Here, we use the \citet{robitaille2017} set of YSO model SEDs and dedicated fitting tool \citep{robitaille2007} to reproduce multi-wavelength photometry of YSO candidates.
\begin{figure}
\begin{center}
\includegraphics[scale=0.7]{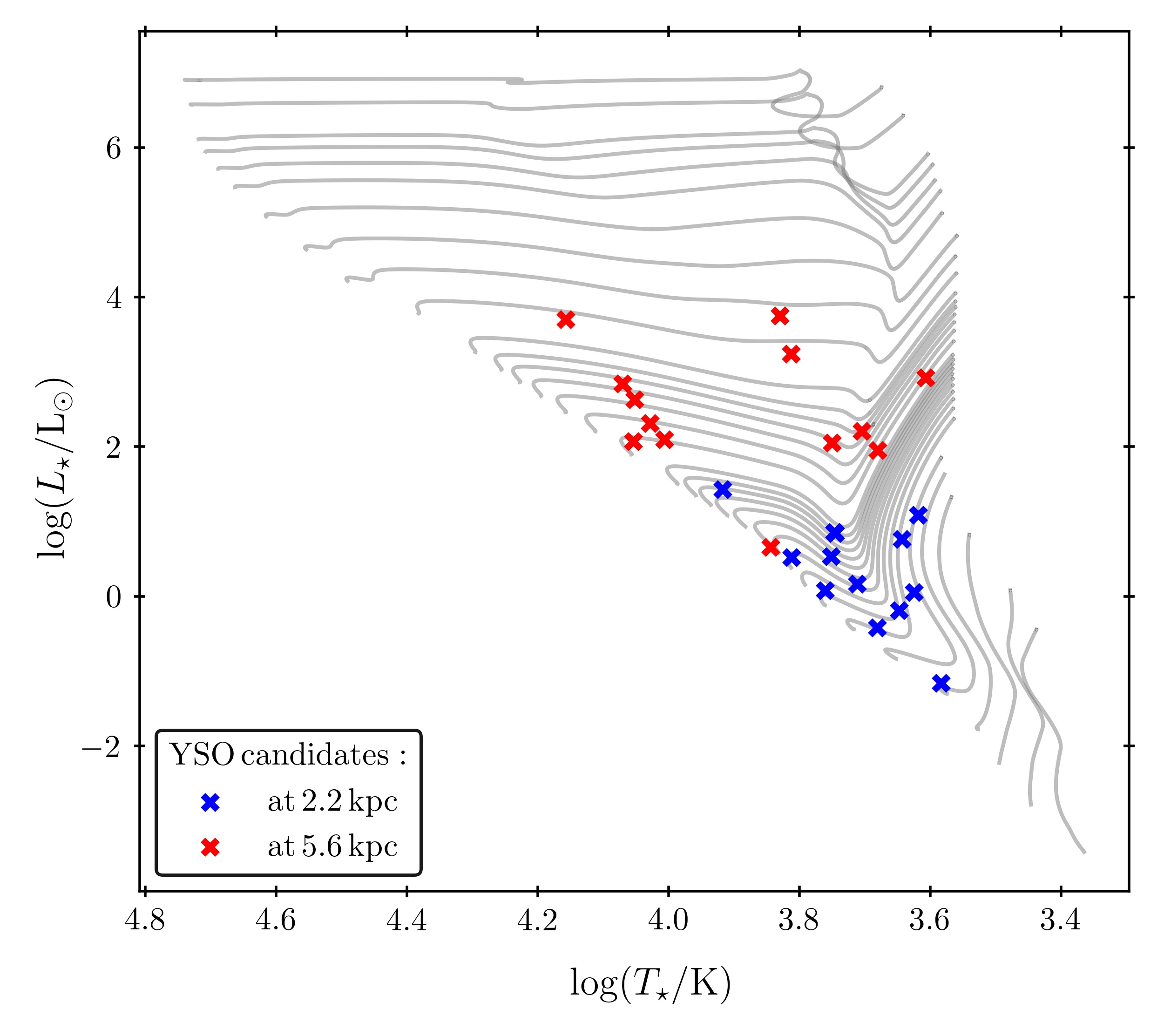}
\end{center}
\caption{HR diagram with YSOs in the IRAS22147 region and the PARSEC evolutionary tracks \citep{bressan2012,chen2014,chen2015,tang2014}. YSOs located at $\sim$2.2 kpc are in blue, and at $\sim5.6$ kpc are in red. Every third track is drawn for clarity.}
\label{fig:hr}
\end{figure}

The \citet{robitaille2017} SED models are divided into 18 sets consisting of a combination of a number of components: star, disc, infalling envelope, bipolar cavities, and an ambient medium (Table \ref{t:modelstats}). The intention is to span a wide range of parameter space, however, as a result many of the models are unphysical. Therefore, we limit our fitting procedure to the models with positions on the Hertzsprung-Russell (HR) diagram that are consistent with the PARSEC evolutionary tracks produced by the revised Padova code \citep{bressan2012, chen2014, chen2015, tang2014}. We excluded the YSO models located outside the coverage of the PARSEC pre-main sequence stage tracks.

In the fitting procedure, we set a 10\% error on the distance and varied the interstellar extinction, $A_V$, in the range from 0 to 40, using the \citet{weingartner2001} extinction law for $R_V=4.0$. We model only the sources which have at least five photometric detections, including at least one at wavelength $\gtrsim 12\,{\rm \mu m}$. In addition, we use only AllWISE 12 and 22\,${\rm \mu m}$ data, since the 3.5 and 4.6\,${\rm \mu m}$ wavebands are similar to those of {\it Spitzer}, but with significantly lower angular resolution (6$"$ in WISE vs. 2$"$ {\it Spitzer}). In the IRAS 22147 region, a total of 26 sources pass these criteria.

In order to find the best fit, the code initially determines the fit with the lowest value of $\chi^2$ and $\chi^2_{\rm best}$. In each of the model sets, the number of fits with the $\chi^2<(\chi^2_{\rm best}+F)$ is counted, where $F$ is a threshold parameter which we set to 3 \citep{sewilo2019}. The best SED fit is the one with the lowest $\chi^2$ from the model set with the largest number of good fits. The above conditions are adopted to ensure the proper selection of the model and limit the risks stemming from the fact that the SED fitting is degenerate, with some of the model sets including 13 parameters.

Table \ref{t:modelstats} gives the statistics of the best-fit SED models for the YSO candidates in the IRAS 22147 region, with Figure\,\ref{seds} showing two example best-fit SEDs. Twenty-three sources are successfully modeled with a star and a passive disk, corresponding to Class II+III sources. The model for one source requires the presence of a protostellar envelope (Class 0), with an additional two features, including envelopes as well as passive disks (Class I).

In order to determine stellar luminosities, masses, and ages for our sample, which are not provided by the \citet{robitaille2017} SED models, we use the following procedure. First, we find the luminosity using the Stefan-Boltzmann law, adopting the stellar radius and effective temperature from the SED fitting. Then, the mass and age are determined from the closest pre-main sequence (PMS) track, found for each model on the HR diagram (see Appendix \ref{sec:seds2}).
\begin{figure*}
\centering

\includegraphics[scale=0.7]{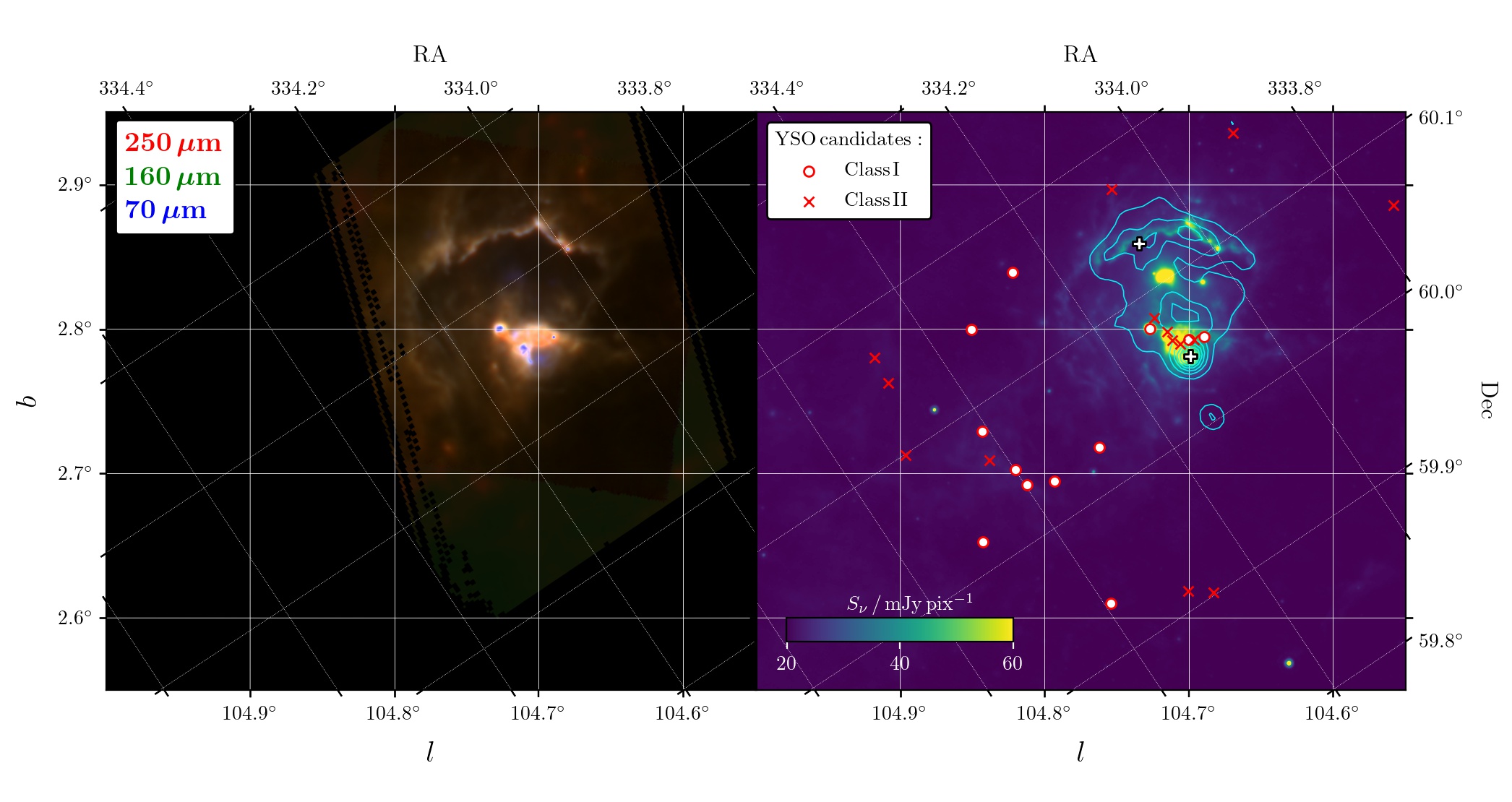}
\caption{Far-IR and radio images of the IRAS 22147 region. \textbf{Left:} Three-color composite image of the IRAS 22147 region combining the \textit{Herschel}/EPoS SPIRE 250 $\mu$m (red), PACS 160 $\mu$m (green), and PACS 70 $\mu$m (blue) mosaics. \textbf{Right:}  \textit{Spitzer}/MIPS 24 $\mu$m image with the VLA 1.4 GHz radio continuum contours overlaid \citep{green1994}. Positions of YSO candidates identified in this work are shown in red symbols, as specified in the legend. White xs show central positions of \ion{H}{ii} regions G104.735+02.859 at 7.8 kpc (position at the arc) and G104.699+02.781 at 6.6 kpc \citep{anderson2015}.}
\label{comp}
\end{figure*}

We adopted the stellar masses only for the sources, for which the age from PMS tracks is consistent with the SED fitting value. We use the YSO lifetimes determined by \citet{dunham2015}, based on a sample of $\sim$3000 YSO candidates from nearby star-forming regions, with Class 0+I YSOs lifetimes of 0.46-0.72 Myr (and duration of 0.335 and 0.665 $\times$ lifetime for Class 0 and I, respectively), flat-spectrum sources lifetimes of 0.30-0.46 Myr and 3 Myr for Class II+III.

Table\,\ref{t:physpar} shows a comparison of the YSO classification using color-color diagrams and the SED modeling, and the corresponding physical parameters. Fully consistent object classification with the two methods is obtained for 15 out of 26 sources including two Class I sources and 13 Class II and III sources. One of the Class II YSOs from the color-color diagrams is re-classified as Class 0 source using SED modelling. The remaining ten sources originally identified as embedded Class I protostars are identified as more evolved Class II objects based on the SED models. Without far-IR data, it is difficult to distinguish Class I YSOs from reddened Class II YSOs \citep{kry12}. Two of these ten sources have far-IR photometry that supports their identification as Class II objects. The nature of the other eight sources is more ambiguous due to the large range of far-IR colors among their best-fitting SED models \citep{ali10}, but they likely lie near the Class I-II transition. Given these uncertainties, we adopted a classification based on color-color diagrams for the discussion of YSO candidates in the IRAS 22147 region.

Figure \ref{fig:hr} shows the Hertzsprung-Russell Diagram with the positions of the YSOs obtained from the SED modeling (Table \ref{t:physpar}). The spread in YSO age is similar to the one found for the Orion Trapezium Cluster, however, at lower luminosities \citep{fang21}. Due to larger distances probed here, the YSOs in the IRAS 22147 region probe higher mass tip of of a full mass distribution. The split in the 5.6 kpc group, with about half the stars at young ages and half quite close to the main sequence, might be an evidence for two populations. Follow-up spectroscopy would be required to confirm such a scenario.

\section{Discussion} 
\label{sec:disc}
\subsection{Star formation in the IRAS 22147 region}
\label{sec:sf}
A multi-wavelength analysis of the IRAS 22147 region confirms an on-going star formation associated with \iras{}, as well as with the nearby IRAS 22159+5948 (Section \ref{sec:res}). Altogether, 26 YSOs have been identified and classified as Class 0, I, or II objects using standard color-color diagrams, the OCSVM machine learning algorithm, and the SED modeling. Out of 26 sources, 13 are spatially associated with the CO 1-0 emission at 5.6 kpc and the dust emission seen in Fig. 1. Those sources are grouped in two compact clusters (Fig. 6), while the remaining YSOs form a more loosely distributed, foreground group at 2.2 kpc. Star formation in the smaller cluster toward IRAS 22159+5948 might have been triggered by the earlier formation of stars near the position of \iras{}, given that third-fourths of YSOs, there are Class I objects based on the color-color diagrams method with regard to third-ninths of objects in the main cluster. 

The main ambiguity in the interpretation of the IRAS 22147 region was due to early observations in the radio regime, showing a steep radio spectrum typical for supernovae remnants \citep{green1994}. However, subsequent measurements at 10.5 GHz\footnote{See a note added in proof in \cite{green1994}} were no longer consistent with the non-thermal emission. Subsequent detection of radio recombination lines with the Green Bank Telescope \ion{H}{ii} Region Discovery Survey \citep{bania2010} and significant infrared emission with WISE \citep{anderson2015} provided strong arguments in favor of the source classification as the \ion{H}{ii} region.

Figure 10 compares the mid-infrared dust emission at 24 $\mu$m with the 1.4 GHz emission from Very Large Array \citep[VLA; ][]{green1994}, along with the positions of \ion{H}{ii} regions from t he WISE Catalog of Galactic \ion{H}{ii} regions \citep{anderson2015}. The radio emission resembles the infrared structures and peaks in the vicinity of the main cluster associated with \iras{} and along the northern arc, suggesting a physical connection. However, according to \cite{anderson2015} the distance to \ion{H}{ii} region G104.735+02.859 (at the arc) is 7.8 kpc and to G104.699+02.781 is 6.6 kpc. The distance of 7.8 kpc is clearly inconsistent with the distance of 5.6 kpc determined using CO 1-0 (Section \ref{sec:res}) and suggests that the radio source is not related to \iras{}. On the contrary, the distance of G104.699+02.781 roughly agrees with the CO measurements and the differences may be due to uncertainties in the adopted model of the Galaxy for the outer Galaxy \citep{konig21}. The velocity of hydrogen radio recombination lines of -57.8 km s$^{-1}$ corresponds to the distance of $5.17\pm0.6$ kpc according to the updated rotational curve from the Bar and Spiral Structure Legacy Survey \citep[BeSSeL; ][]{reid14}. The distance to \iras{} using this slightly updated rotation curve (w.r.t. the one we use in Section \ref{sec:res}) is 5.33$\pm0.64$ and in excellent agreement with the distance of 5.29$\pm0.65$ obtained as part of the EPoS survey \citep{ragan2012}. Thus, the \ion{H}{ii} region G104.699+02.781 is indeed located in the close vicinity of \iras{} and might have been a driving force triggering lower-mass star formation in the region. 

Several YSOs in the IRAS 22147 region are associated with gas at a distance of 2.2 kpc, illustrating the problem with cloud confusion during the interpretation of infrared observations in the outer Galaxy. \cite{kerton2003} developed a method of matching IRAS sources with proper 
CO clouds based on probability considerations using random lines of sight. Among all IRAS sources covered by the FCRAO survey of the outer Galaxy, they found multiple CO clouds along the line of sight toward 56.5\% of IRAS sources and 78\% of IRAS sources studied with pointed CO observations by \cite{wb89}. These authors also found a single distance to \iras{} consistent with our study, and correctly assigned a higher probability of a similar distance to \iras{}. Clearly, the interpretation of star forming regions in the outer Galaxy require a multi-wavelength approach and a combination of state-of-the art methods to properly identify YSOs and the properties of their host clouds and clumps.

\subsection{Utility of the machine learning algorithms}
\label{sec:algo}

Machine learning algorithms are now widely used in various astrophysical applications and clearly provide a useful tool also for the identification of YSOs and their clusters in the outer Galaxy.

In Section \ref{sec:ml}, we used the OCSVM algorithm to identify YSOs in the mid-infrared photometry from WISE and obtained 23 new YSO candidates, which accounts for 47\% of detections using all adopted methods. Out of 23 objects, 18 are new, unique detections that could not be identified based on the color-color criteria used by
\cite{Winston2019}. We verified the location of the YSOs selected by the algorithm on the color-color diagrams and found a confirmation of these findings by comparisons with spectroscopically-confirmed YSOs in the LMC \citep{jones17}. The spectroscopy of all sources in the IR or submillimeter would be necessary to double-check their status, similar to the LMC efforts, but given the vastness of the outer Galaxy the OCSVM algorithm offers a compelling alternative. 

The choice of the machine learning algorithm (and the adopted parameters) influences the final results and shall be well-matched with the specifics of the problem to be solved. In Section \ref{sec:clust}, we used the agglomerative hierarchical clustering algorithm to identify YSO clusters in \iras{} \citep{everitt11}, whereas \citet{Winston2019} selected the density-based spatial clustering and application with noise method \citep[DBSCAN, ][]{ester1996}, which is a density-based clustering algorithm that can be used to identify any cluster shape in the data set containing noise and outliers. It identifies dense regions, which can be measured by the number of objects close to a given point. When applying the DBSCAN algorithm to our YSO sets within the IRAS 22147 region, we are able to recover only a single cluster located at 5.6 kpc. Clearly, DBSCAN works better with large data-sets and it does not assign every point to a cluster. As opposed to the agglomerative hierarchical clustering algorithm, DBSCAN will not partition the data but, rather, will extract the "dense" clusters and treat sparse background as noise. For this reason, in the case of a relatively small IRAS 22147 region, hierarchical clustering is a more reliable and effective approach.

Various machine learning algorithms have been tested by other authors to identify YSOs on a statistical basis using large-sky surveys. For example, the SVM algorithm used in our analysis has been adopted to an all-sky selection of YSOs from the AllWISE catalog leading to the identification of more than 90\% YSOs from older catalogs \citep{marton2016}. Additional algorithms such as random forests and neural networks were used to refine earlier identifications and benefit from distances from the Gaia DR2 catalog \citep{marton2019}. As a result, 1 129 295 YSO candidates have been identified in the entire sky, including 889 in the IRAS 22147 region alone (assuming probability above 0.6 and the matching radius of 2''). Thus, the results do not seem to be sufficiently selective, which is further evidenced by a significant number of sources at high latitudes. 

A recent study of Herbig Ae/Be and classical Be stars with algorithms using neural networks provides more realistic results, likely due to narrowing down the source characteristics and more extended source information \citep{vioque2020}. Certainly, less evolved, Class 0 or I sources form the largest challenge for such studies given the relatively poor angular resolution of mid- and far-IR data. Thus, a combined analysis of machine learning algorithms with standard methods involving color-color diagrams, dust and gas emission, and CO distances are the most optimal alternatives to spectroscopic surveys. 

\section{Conclusions} 
\label{sec:concs}
The SMOG survey and complementary far-IR and CO data allow us to identify and characterize a population of YSO candidates in the IRAS 22147 Outer Galaxy region. A combination of standard color-color diagram method and the machine learning techniques, as well as information on gas and dust spatial distribution and SED modeling are necessary to avoid cloud confusion and properly determine the census of YSOs in specific star forming regions. The main conclusions of our study are as follows.  

\begin{enumerate} 

\item{The IRAS 22147 region is associated with two distinct star forming regions located at $\sim$2.2 and 5.6 kpc. The compact, central embedded cluster is the more distant region physically connected with the northern arc of material and a south-west cluster associated with IRAS 22159+5948. The region at 2.2 kpc shows a more dispersed population of YSOs.}

\item{Using standard methods based on photometry and color-color cuts, we identify 13 Class I and 13 Class II sources, including 6 Class I and 7 Class II associated with the embedded cluster at 5.6 kpc associated with \iras{}. A smaller cluster at 5.6 kpc, which is co-spatial with IRAS 22159+5948, contains 3 Class I and 1 Class II YSO candidates, suggesting a later launch of star formation (w.r.t. the IRAS 22147 region). }

\item{Two Class I sources and 21 Class II sources are identified using the SVM learning algorithm which uses mid-infrared AllWISE catalog. New boundaries in the color-color diagrams based on AllWISE filters are needed to identify YSOs in future studies.}

\item{Five clusters containing 16 YSO candidates are identified in the IRAS 22147 region. Two clusters at 5.6 kpc are confirmed by the spatial extent of CO 1-0 and dust emission. Three clusters at 2.2 kpc are also co-spatial with available CO 1-0, but form a loosely connected population which might as well be regarded as a single, large cluster. The agglomerative hierarchical clustering algorithm used in the analysis successfully resolved clusters at 5.6 kpc, unlike the commonly used DBSCAN, which seems to be better suited to large datasets.}

\item{Physical parameters for 26 YSOs were obtained from the YSO SED model fitting \citep{robitaille2017}. The modeling confirms their status as YSOs and provides complementary measure to the classification obtained with color-color and machine-learning methods. However, the classification is strongly affected by the lack of high angular-resolution far-IR photometry.}

\end{enumerate}

Multi-wavelength datasets that include infrared wavelengths are necessary for the proper identification and classification of YSOs, in particular those in the deeply-embedded stages. High angular-resolution submillimeter spectroscopy of Class 0 and I sources are needed to investigate outflow activity and molecular abundances in the outer Galaxy. The latter is a necessary step toward the determination of the gas temperatures, densities, and UV radiation field strengths, as well as improving our understanding of the impact of metallicity on star formation.
 
\begin{acknowledgements}  The authors
would like to thank the anonymous referee for the comments and recommendations which helped to improve this manuscript. This work was supported by the Polish National Science Center grants 2014/15/B/ST9/02111 and 2016/21/D/ST9/01098. AK and MK acknowledge support from the First TEAM grant of the Foundation for Polish Science No. POIR.04.04.00-00-5D21/18-00. This article has been supported by the Polish National Agency for Academic Exchange under Grant No. PPI/APM/2018/1/00036/U/001. The material is partially based upon work supported by NASA under award number 80GSFC21M0002. The research presented in this paper has used data from the Canadian Galactic Plane Survey, a Canadian project with international partners, supported by the Natural Sciences and Engineering Research Council. The Dominion Radio Astrophysical Observatory is operated as a national facility by the National Research Council of Canada. The Five College Radio Astronomy Observatory CO Survey of the Outer Galaxy was supported by NSF grant AST 94-20159.
\end{acknowledgements}

\bibliographystyle{aa} 
\bibliography{biblio_IRAS22147.bib}

\begin{appendix}
\section{Kinematic distances versus the Gaia measurements}
\label{sec:gaia}

\begin{table*}[ht!] 
\caption{Distances to YSOs based on the CO emission and the Gaia catalog \citep{gaia21} \label{t:gaia}}
\centering                        
\begin{tabular}{rcccccccccc}
\hline \hline                 
Nr & IRAC Designation\tablefootmark{a} & RA & DEC & Separation\tablefootmark{b} & $D_\mathrm{kin}$\tablefootmark{c} & 
\multicolumn{3}{c}{$D_\mathrm{Gaia}$\tablefootmark{d} (pc)}  \\  \cline{7-9} 
~ & `SSTSMOGA'  & (deg) & (deg) & (arc sec) & (pc) & 50$^{\mathrm{th}}$ & 14$^{\mathrm{th}}$ & 86$^{\mathrm{th}}$ \\ 
\hline
1       &       \textbf{G104.6313+02.6211}      &       334.176540      &       59.876274       &       0.60    &       2197    &       224     &       218     &       231     \\      
2       &       G104.6829+02.6173       &       334.266213      &       59.901773       &       0.12    &       2197    &       1481    &       1227    &       1968    \\      
\hline
3       &       G104.7023+02.8142       &       334.079104      &       60.076155       &       0.11    &       5628    &       6245    &       4413    &       9334    \\      
4       &       G104.7061+02.7861       &       334.116800      &       60.054850       &       0.35    &       5628    &       8687    &       4425    &       10873   \\      
5       &       \textbf{G104.7062+02.7895}      &       334.113200      &       60.057750       &       0.15    &       5628    &       1179    &       1157    &       1200    \\      
6       &       G104.7153+02.7979       &       334.118870      &       60.069817       &       0.11    &       5628    &       9531    &       5675    &       15687   \\      
7       &       G104.7241+02.8075       &       334.122770      &       60.082638       &       0.09    &       5628    &       7498    &       4365    &       9265    \\      
\hline
8       &       \textbf{G104.7418+02.5949}      &       334.388550      &       59.915745       &       0.07    &       2215    &       5354    &       4070    &       6247    \\      
9       &       \textbf{G104.7538+02.8964}      &       334.072570      &       60.172980       &       0.17    &       2208    &       506     &       454     &       577     \\      
10      &       \textbf{G104.8429+02.6522}      &       334.493160      &       60.019398       &       1.05    &       2230    &       6691    &       4698    &       9651    \\      
11      &       \textbf{G104.9084+02.7625}      &       334.480220      &       60.147484       &       0.02    &       2230    &       3839    &       3109    &       4469    \\      
12      &       G104.9180+02.7798       &       334.477020      &       60.167213       &       0.18    &       2223    &       1673    &       1440    &       2567    \\      
\hline 
\end{tabular}
\begin{flushleft}
\tablefoot{
\tablefoottext{a}{Sources with the differences in the distances calculated using CO and Gaia exceeding the Gaia uncertainties are in bold.}
\tablefoottext{b}{A separation between the YSO and the closest Gaia target from the \cite{gaia21} catalog.}
\tablefoottext{c}{A kinematic distance from the CO emission (Section \ref{sec:dist}).}
\tablefoottext{d}{The three quantiles of a distance distribution from \cite{gaia21}: the median (the 50th percentile), the 14th, and 86th percentiles.}
}
\end{flushleft}
\end{table*}

\begin{figure}[ht!]
\centering
\includegraphics[scale=0.75]{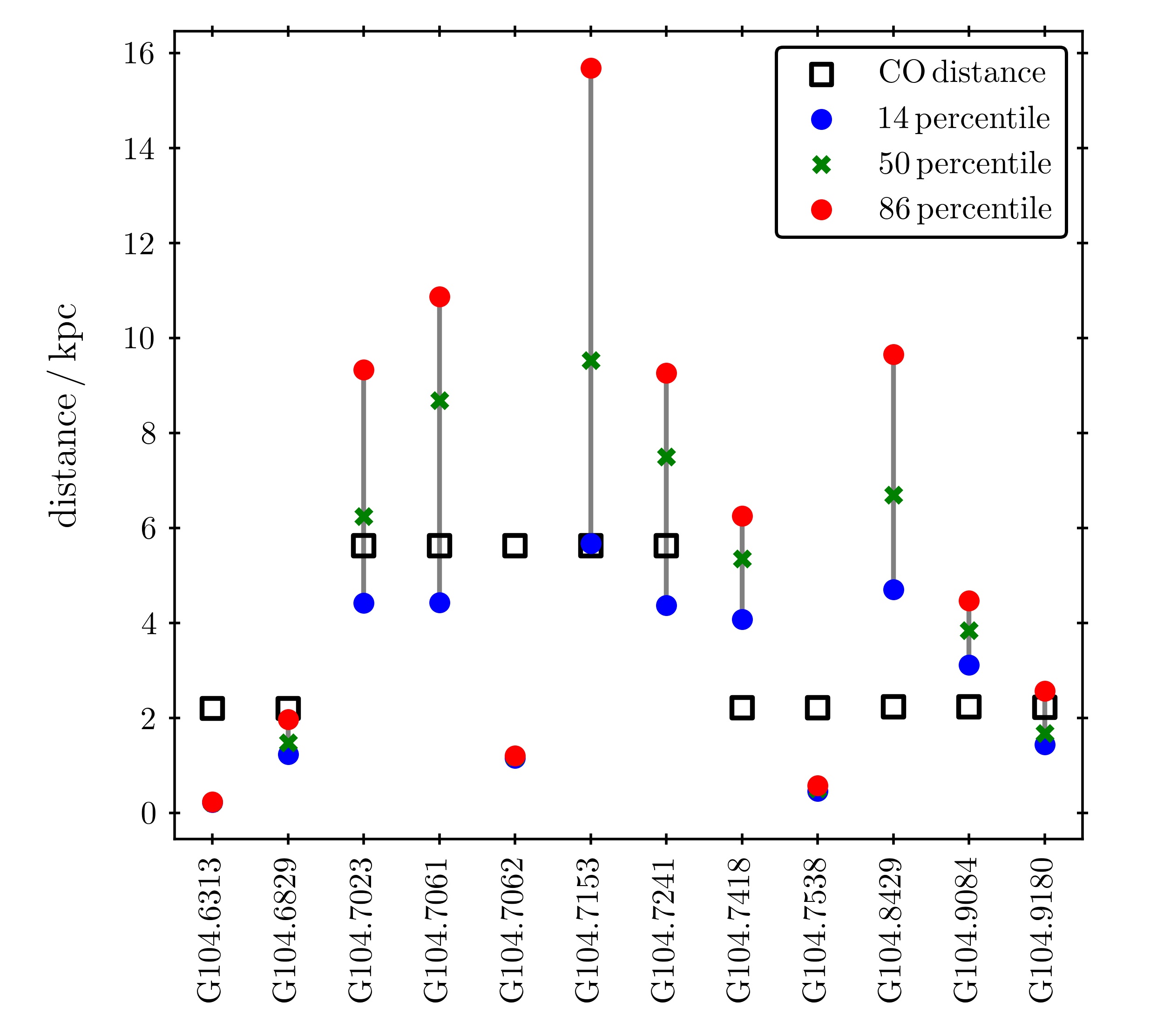}
\caption{Comparison of the kinematic distances of YSOs in the IRAS 22147 region obtained using CO observations and the distances from Gaia \citep{gaia21}. Black empty squares show the kinematic distance obtained using CO emission (Section \ref{sec:dist}). Blue dots, green crosses, and red dots show the 14th, 50th (median), and the 86th percentile of photogeometric distance distribution. The full names of YSOs and their equatorial coordinates are provided in Table \ref{t:gaia}.}
\label{gaia}
\end{figure}

\begin{figure*}[ht!]
\centering
\includegraphics[scale=0.6]{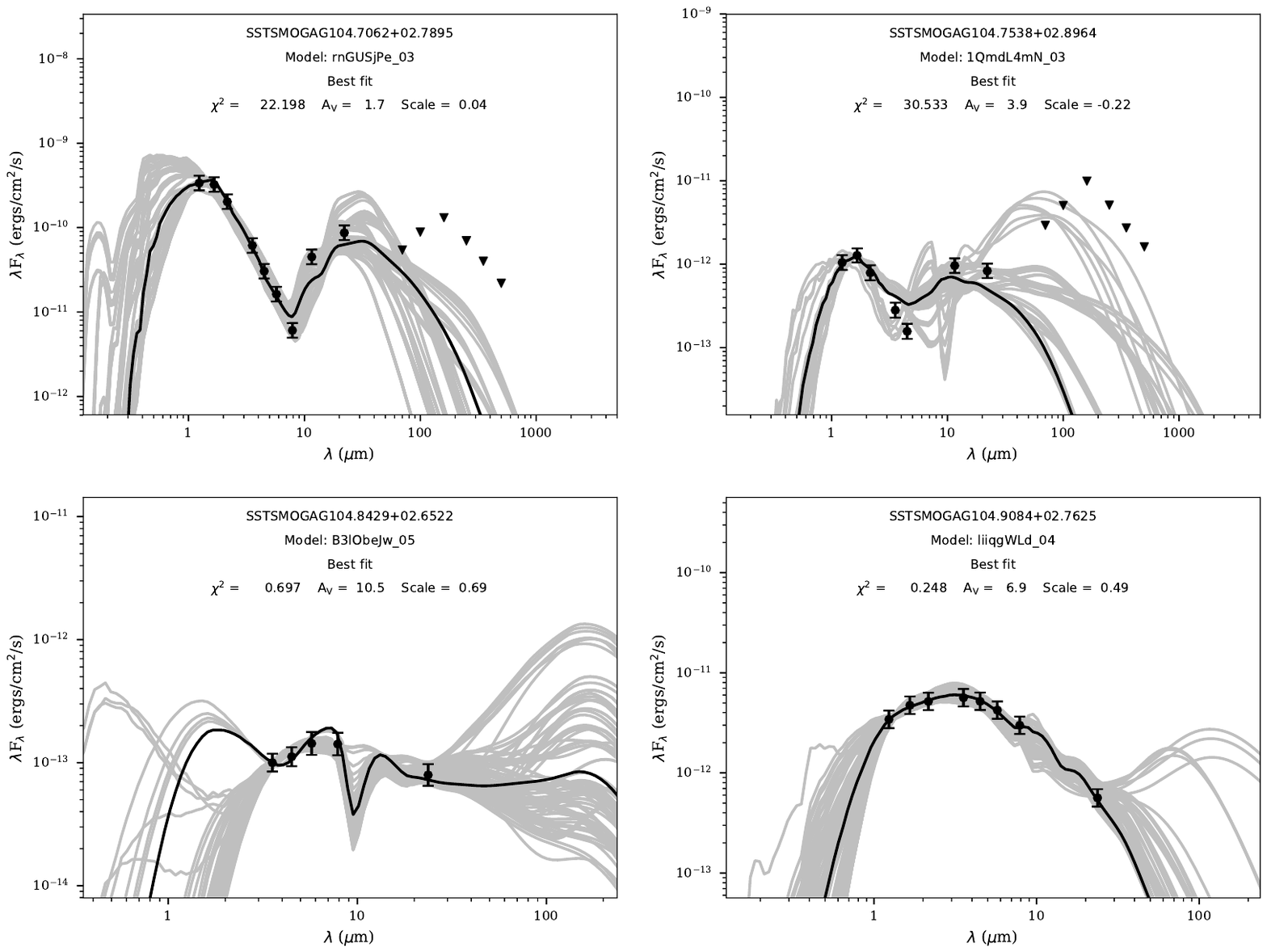}
\caption{SEDs of candidate YSOs with distances adopted from the Gaia DR3 catalog (see Table \ref{t:physgaia}), see also Fig. \ref{seds}.}
\label{gaia_sed}
\end{figure*}

\begin{table*}[ht!]
\caption{Physical Parameters for a Subset of YSO Candidates in the IRAS 22147 region with distances from the Gaia DR3 catalog \label{t:physgaia}}
\centering                        
\begin{tabular}{rcccccccccc}
\hline \hline                 
Nr & IRAC Designation & Model & Dist.$^b$ & $R_{\rm \star}$ & $T_{\rm \star}$ & $L_{\rm \star}$ & $M_{\rm \star}$ & Age \\
 & `SSTSMOGA'        & Set$^a$ & (kpc) & ($R_{\odot}$)  & ($T_{\odot}$)   & ($L_{\odot}$)   & ($M_{\odot}$)       & (Myr)   \\
\hline
1 & G104.7062+02.7895 & sp--h-i & 1.2  & $9.3$ & $4500$ & $31.8$ & 1.3 & 0.3 \\ 
2 & G104.7538+02.8964 & sp--h-i & 0.5  & $0.5$ & $3440$ & $0.04$ & -- & -- \\ 
3 & G104.8429+02.6522 & sp--smi & 6.7 & $1.7$ & $6220$ & $3.7$ & -- & -- \\ 
4 & G104.9084+02.7625 & sp--s-i & 3.8 & $2.9$ & $7210$ & $20.7$ & -- & -- \\ 
\hline 
\end{tabular}
\begin{flushleft}
$^a$See Tables~\ref{t:modelstats} and \ref{t:physpar} for the meaning of characters in the model set names, and the remarks to the physical quantities, respectively.\\
$^b$ Distances taken from \cite{gaia21}. \\
\end{flushleft}
\end{table*}

We compared kinematic distances obtained using CO (Section \ref{sec:dist}) with the Gaia Early Data Release 3 catalog \citep{brown21}. We cross-correlated 49 YSOs in the IRAS 22147 region with the catalog by \cite{gaia21}, which provides photogeometric distances for 1.35 billion Gaia targets. The distances are determined using a probabilistic approach involving information about the colors and apparent magnitudes of stars. 

Twelve Gaia targets are found within $\sim$1 arc sec separation from YSOs in the IRAS 22147 region ($\sim$25\% of all YSOs). The remaining YSOs are located more than $\sim$2.5 arcsec away from the Gaia targets, and are unlikely to be their true counterparts. Table \ref{t:gaia} shows the comparison of the distances obtained toward the matched sources using the CO emission (Section \ref{sec:dist}) and the photogeometric distances from Gaia \citep{gaia21}. Figure \ref{gaia} shows the same information in a graphical way, illustrating the uncertainties of the Gaia measurements. The uncertainties of the kinematic distances from CO depend on the accuracy of the rotational curves and proper motions of the sources, and may be probably as high as 50\% of the quoted values.

Among the YSOs located at $\sim$5.6 kpc, the kinematic distances of four out of five objects are consistent with the Gaia distances within uncertainties; i is  G104.7062 tat appearsto be located at $\sim2.2$ kpc. The discrepancies are more severe for objects with kinematic distances of $\sim$2.2 kpc, with only two out of seven objects showing values consistent with Gaia. In case of G104.7418 and G104.8429, the Gaia distances suggest the position of the objects in the more distant cloud (at $\sim5.6$ kpc). On the contrary, G104.6313 and G104.7538 are located at even more nearby star-forming regions.

The discrepancies between kinematic distances and Gaia measurements point to the problem with the automatic selection of the best CO clump in case of CO lines at multiple velocities toward the same sources. In case of G104.7062, G104.6313, and G104.7538, the distances from Gaia are very precise and likely more reliable than the kinematic ones, due to uncertainties in using the rotational curve for the most nearby sources. Thus, we adopted the median distances from Gaia for the subsequent analysis of those objects.

\begin{figure}[ht!]
\centering
\includegraphics[scale=0.75]{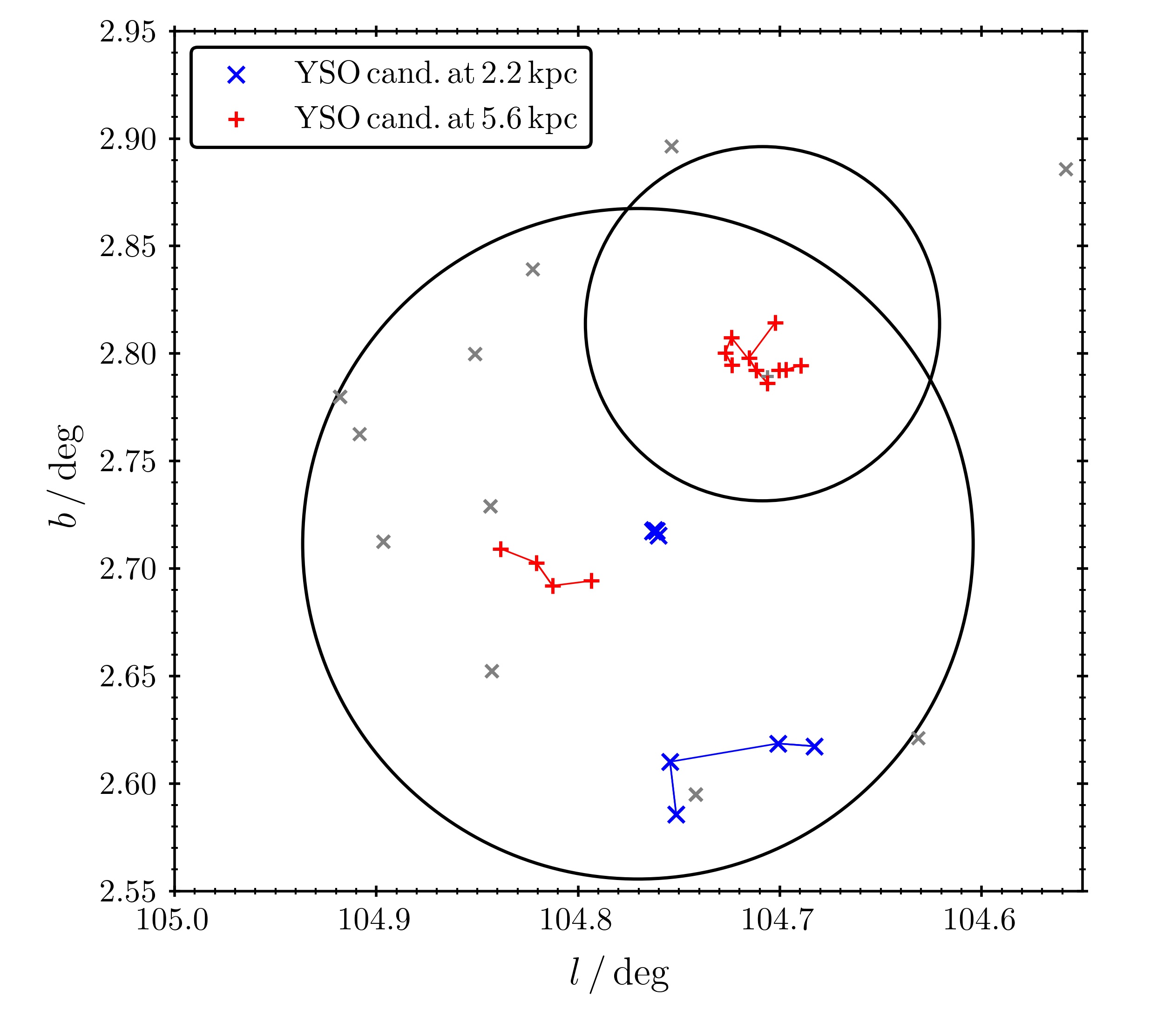}
\caption{Minimum spanning tree for YSO candidates in the IRAS 22147 region adopting Gaia distances for G104.7062, G104.6313, and G104.7538 (shown as gray x's), see also Fig. \ref{mst}. }
\label{gaia:clust}
\end{figure}

Seven YSOs with Gaia distances passed our criteria for the SED modelling (Table \ref{t:physpar}). Three sources show sufficient agreement with the kinematic distances, which are adopted for the modeling (G104.6829, G104.7153, G104.9180). For the remaining four sources, we used Gaia distances to obtain new YSO SED models (G104.7062. G104.7538, G104.8429, and G104.9084). 

Table \ref{t:physgaia} shows the physical parameters of YSOs in the IRAS 22147 region obtained from the best-fit \cite{robitaille2017} SED models and the Gaia distances from \citealt{gaia21} (Table \ref{t:gaia}). An decrease of almost two orders of magnitude  in source luminosity is seen for G104.7062 (from 1750 to 32 L$_{\odot}$) and G104.7538 (from 3 to 0.04 L$_{\odot}$) due to a factor of 4-5 smaller distances to these sources found by Gaia, as well as a slightly different sets of YSO models fitted (Table \ref{t:modelstats}). Similarly, the luminosity of G104.8429 increases from 0.1 to 3.7 L$_{\odot}$ due to a larger distance to the source from Gaia by factor of 3. Finally, a similar source luminosity of G104.9084 (20.7 L$_{\odot}$) is obtained using the Gaia distance of 3.8 kpc and the kinematic distance of 2.2 kpc (27 L$_{\odot}$). The classification of the sources is not altered except G104.7062, ages of which are consistent with Class 0/I protostars. Figure \ref{gaia_sed} shows the SEDs of YSOs and the best-fit \cite{robitaille2017} models (see also Appendix \ref{sec:seds2}).

Figure \ref{gaia:clust} shows the impact of adopting the Gaia distances (Table \ref{t:gaia}) on the clustering analysis. Among six sources with the significant differences between Gaia and CO distances, only three sources are cluster members when CO kinematic distances are assumed (G104.7062, G104.7418, and G104.9084; see also Section \ref{sec:clust}). Adopting the distance of 1.2 kpc for G104.7062 would result in the removal of the source from the cluster at 5.6 kpc; the ratio of Class II to Class I objects would decrease from 2.7 to 2.3, but the cluster center remains almost the same at l,b=(104.100$^{\mathrm{o}}$, 2.797$^{\mathrm{o}}$). Similarly, adopting a distance of 5.4 kpc for G104.7418 would remove it from the originally five-member cluster at 2.2 kpc; the ratio of Class II to Class I objects would decrease from 4.0 to 3.0 and the cluster center moves to (104.722$^{\mathrm{o}}$, 2.608$^{\mathrm{o}}$). Finally, G104.9084 is a member of a small, two-member cluster at 2.2 kpc, and adopting the new Gaia distance of 3.8 kpc would result in the cancellation of the cluster identification.

We note that the Gaia distances and their impact on the clustering does not seem to be consistent with the distribution of molecular gas in the IRAS 22147 region (Fig. \ref{fig:co}). G104.7062 is located in the middle of the cluster at 5.6 kpc, where significant amount of molecular gas is present. The CO spectra at those positions do not contain any additional CO lines at velocities corresponding to cloud at smaller distances. A similar situation occurs at the southern cluster with G104.7418, which is located in the central part of the 2.2 kpc cluster, with no CO emission corresponding to larger distances. We conclude that the Gaia distances may not provide the improvement over the kinematic distances in case of star-forming regions. 

\section{Photometry of YSO candidates in the IRAS 22147 region}
\label{sec:phottab}

Tables \ref{tab:photo} and \ref{tab:photo2} show the multi-wavelength photometry for 35 YSOs in the IRAS 22147 region identified using the color-color diagrams and the SVM learning algorithm, for which the distances have been assigned (see Section \ref{sec:res}). The table includes nine sources without 12 $\mu$m photometry for which the SED modeling was not performed.

Table \ref{tab:photo_no_d} shows the photometry of the remaining 14 YSOs identified in the IRAS22147 region, for which the distances are not available. The SEDs of these
sources could not have been modeled with the \cite{robitaille2017} YSO models (Section \ref{sec:seds}).

\section{Comparison of the YSO classification with the literature results}
\label{sec:class}

Table \ref{t:winston} shows a source-by-source comparison of the YSO classification for 24 sources that were identified here and in \cite{Winston2019} (see Section \ref{sec:complet}). For completeness, we include also the classification based on color-color diagrams, where we did not require the distance determination (column \lq CC (no dist.)'). This method is not optimal, as we could not properly isolate AGN candidates, but allows us to double-check the classification obtained using the machine learning techniques (column \lq ML').                 

A comparison of the columns labeled "ML" and " MIPS" shows a very good agreement between our work and \cite{Winston2019}, with 14 Class I and 10 Class II sources identified here, and 11 Class I and 13 Class II sources cataloged in \cite{Winston2019}. Six sources have different classification: four are classified as Class I here and as Class II in  \cite{Winston2019}, and two are classified here as Class II (ML) and as Class I in \cite{Winston2019}. One additional source, G104.8383+02.7091, is classified as Class II in our catalog and in \cite{Winston2019} using IRAC and MIPS bands, but appears as Class I when only IRAC bands are used in both works; we consider it as a Class II source in this paper. Overall, 75\% of the YSO candidates in both works share the same classification. 

\begin{sidewaystable*}[ph!]
\begin{center}
\caption{Near- and mid-infrared photometry of YSO candidates in the \iras{} field. The columns represent the 2MASS $JHK_s$, {\it Spitzer} IRAC, and AllWISE 12\,${\rm \mu m}$ bands. Units are mJy and the upper limits are $3\sigma$.\label{tab:photo}}
\setlength{\tabcolsep}{2 mm}
\begin{small}
\begin{tabular}{cccccccccc}
\hline
\hline
Nr & IRAC Designation & \multicolumn{3}{c}{2MASS} & \multicolumn{4}{c}{IRAC} & WISE \\
~ & SSTSMOGA & $J$ & $H$ & $K_s$ & 3.6 $\mu$m & 4.5 $\mu$m & 5.8 $\mu$m & 8.0 $\mu$m & 12 $\mu$m \\
\hline
1 & G104.5582+02.8857 & -- & -- & -- & $\phantom{1}1.29 \pm 0.26\phantom{1}$ & $\phantom{10}1.51 \pm 0.30\phantom{1}$ & $\phantom{10}1.61 \pm 0.32\phantom{1}$ & $\phantom{10}1.69 \pm 0.34\phantom{1}$ & -- \\
2 & G104.6313+02.6211 & $\phantom{10}1.04 \pm 0.21\phantom{1}$ & $\phantom{10}0.93 \pm 0.19\phantom{1}$ & $\phantom{10}0.90 \pm 0.18\phantom{1}$ & $\phantom{1}0.50 \pm 0.10\phantom{1}$ & $\phantom{10}0.36 \pm 0.07\phantom{1}$ & $\phantom{10}0.26 \pm 0.05\phantom{1}$ & -- & -- \\
3 & G104.6697+02.9357 & -- & $\phantom{10}0.37 \pm 0.07\phantom{1}$ & $\phantom{10}0.61 \pm 0.12\phantom{1}$ & -- & $\phantom{10}1.52 \pm 0.30\phantom{1}$ & -- & $\phantom{10}2.21 \pm 0.44\phantom{1}$ & $\phantom{10}1.62 \pm 0.32\phantom{10}$ \\
4 & G104.6829+02.6173 & $\phantom{10}0.80 \pm 0.16\phantom{1}$ & $\phantom{10}1.59 \pm 0.32\phantom{1}$ & $\phantom{10}1.80 \pm 0.36\phantom{1}$ & $\phantom{1}1.60 \pm 0.32\phantom{1}$ & $\phantom{10}1.50 \pm 0.30\phantom{1}$ & $\phantom{10}1.32 \pm 0.26\phantom{1}$ & $\phantom{10}1.65 \pm 0.33\phantom{1}$ & -- \\
5 & G104.6895+02.7945 & -- & $\phantom{10}1.26 \pm 0.25\phantom{1}$ & $\phantom{10}8.68 \pm 1.74\phantom{1}$ & $63.09 \pm 12.62$ & $128.00 \pm 25.60$ & $230.30 \pm 46.06$ & $300.60 \pm 60.12$ & $484.29 \pm 96.86\phantom{1}$ \\
6 & G104.6968+02.7925 & -- & -- & -- & $\phantom{1}0.79 \pm 0.16\phantom{1}$ & $\phantom{10}1.59 \pm 0.32\phantom{1}$ & $\phantom{10}2.63 \pm 0.53\phantom{1}$ & -- & $\phantom{1}57.43 \pm 11.49\phantom{1}$ \\
7 & G104.7003+02.7923 & -- & -- & -- & $\phantom{1}2.01 \pm 0.40\phantom{1}$ & $\phantom{10}3.73 \pm 0.75\phantom{1}$ & $\phantom{10}5.33 \pm 1.07\phantom{1}$ & $\phantom{10}5.60 \pm 1.12\phantom{1}$ & -- \\
8 & G104.7009+02.6186 & -- & -- & -- & $\phantom{1}0.14 \pm 0.03\phantom{1}$ & $\phantom{10}0.17 \pm 0.03\phantom{1}$ & $\phantom{10}0.19 \pm 0.04\phantom{1}$ & $\phantom{10}0.25 \pm 0.05\phantom{1}$ & -- \\
9 & G104.7023+02.8142 & $\phantom{10}1.46 \pm 0.29\phantom{1}$ & $\phantom{10}2.00 \pm 0.40\phantom{1}$ & $\phantom{10}2.11 \pm 0.42\phantom{1}$ & $\phantom{1}2.38 \pm 0.48\phantom{1}$ & $\phantom{10}2.62 \pm 0.52\phantom{1}$ & $\phantom{10}2.36 \pm 0.47\phantom{1}$ & $\phantom{10}2.19 \pm 0.44\phantom{1}$ & $<0.46$ \\
10 & G104.7061+02.7861 & $\phantom{10}1.31 \pm 0.26\phantom{1}$ & $\phantom{10}2.47 \pm 0.49\phantom{1}$ & $\phantom{10}2.90 \pm 0.58\phantom{1}$ & $\phantom{1}2.56 \pm 0.51\phantom{1}$ & $\phantom{10}2.35 \pm 0.47\phantom{1}$ & -- & -- & -- \\
11 & G104.7062+02.7895 & $141.80 \pm 28.36$ & $183.10 \pm 36.62$ & $149.00 \pm 29.80$ & $74.19 \pm 14.84$ & $\phantom{1}46.57 \pm 9.31\phantom{1}$ & $\phantom{1}31.70 \pm 6.34\phantom{1}$ & $\phantom{1}16.24 \pm 3.25\phantom{1}$ & $176.97 \pm 35.39\phantom{1}$ \\
12 & G104.7118+02.7921 & $\phantom{10}1.28 \pm 0.26\phantom{1}$ & $\phantom{1}14.91 \pm 2.98\phantom{1}$ & $\phantom{1}42.53 \pm 8.51\phantom{1}$ & $56.40 \pm 11.28$ & $\phantom{1}39.40 \pm 7.88\phantom{1}$ & -- & $\phantom{1}18.44 \pm 3.69\phantom{1}$ & $202.63 \pm 40.53\phantom{1}$ \\
13 & G104.7153+02.7979 & $\phantom{10}1.71 \pm 0.34\phantom{1}$ & $\phantom{10}2.56 \pm 0.51\phantom{1}$ & $\phantom{10}3.10 \pm 0.62\phantom{1}$ & $\phantom{1}1.94 \pm 0.39\phantom{1}$ & $\phantom{10}1.42 \pm 0.28\phantom{1}$ & -- & -- & $\phantom{1}87.72 \pm 17.54\phantom{1}$ \\
14 & G104.7237+02.7946 & -- & $\phantom{10}0.68 \pm 0.14\phantom{1}$ & $\phantom{10}1.15 \pm 0.23\phantom{1}$ & $\phantom{1}1.41 \pm 0.28\phantom{1}$ & $\phantom{10}1.67 \pm 0.33\phantom{1}$ & $\phantom{10}1.57 \pm 0.32\phantom{1}$ & $\phantom{10}1.42 \pm 0.28\phantom{1}$ & -- \\
15 & G104.7241+02.8075 & $\phantom{10}2.36 \pm 0.47\phantom{1}$ & $\phantom{10}2.92 \pm 0.58\phantom{1}$ & -- & $\phantom{1}2.27 \pm 0.45\phantom{1}$ & $\phantom{10}1.88 \pm 0.38\phantom{1}$ & $\phantom{10}1.29 \pm 0.26\phantom{1}$ & -- & $\phantom{1}17.01 \pm 3.40\phantom{10}$ \\
16 & G104.7270+02.8003 & $\phantom{10}1.51 \pm 0.30\phantom{1}$ & $\phantom{10}5.59 \pm 1.12\phantom{1}$ & $\phantom{1}18.63 \pm 3.73\phantom{1}$ & $89.42 \pm 17.88$ & $142.80 \pm 28.56$ & $196.20 \pm 39.24$ & $348.60 \pm 69.72$ & $900.95 \pm 180.19$ \\
17 & G104.7418+02.5949 & $\phantom{10}3.96 \pm 0.79\phantom{1}$ & $\phantom{10}6.47 \pm 1.29\phantom{1}$ & $\phantom{10}5.78 \pm 1.16\phantom{1}$ & $\phantom{1}3.12 \pm 0.62\phantom{1}$ & $\phantom{10}3.27 \pm 0.66\phantom{1}$ & $\phantom{10}1.30 \pm 0.26\phantom{1}$ & -- & $<0.24$ \\
18 & G104.7513+02.5857 & -- & -- & -- & $\phantom{1}0.23 \pm 0.05\phantom{1}$ & $\phantom{10}0.22 \pm 0.04\phantom{1}$ & $\phantom{10}0.16 \pm 0.03\phantom{1}$ & $\phantom{10}0.20 \pm 0.04\phantom{1}$ & $<0.34$ \\
19 & G104.7538+02.8964 & $\phantom{10}0.44 \pm 0.09\phantom{1}$ & $\phantom{10}0.72 \pm 0.14\phantom{1}$ & $\phantom{10}0.58 \pm 0.12\phantom{1}$ & $\phantom{1}0.34 \pm 0.07\phantom{1}$ & $\phantom{10}0.24 \pm 0.05\phantom{1}$ & -- & -- & $\phantom{10}3.78 \pm 0.76\phantom{10}$ \\
20 & G104.7544+02.6101 & -- & -- & -- & $\phantom{1}2.10 \pm 0.42\phantom{1}$ & $\phantom{10}3.19 \pm 0.64\phantom{1}$ & $\phantom{10}4.09 \pm 0.82\phantom{1}$ & $\phantom{10}4.12 \pm 0.83\phantom{1}$ & $\phantom{10}4.00 \pm 0.80\phantom{10}$ \\
21 & G104.7601+02.7154 & -- & -- & -- & $\phantom{1}0.42 \pm 0.08\phantom{1}$ & $\phantom{10}0.49 \pm 0.10\phantom{1}$ & $\phantom{10}0.62 \pm 0.12\phantom{1}$ & $\phantom{10}0.79 \pm 0.16\phantom{1}$ & -- \\
22 & G104.7611+02.7176 & -- & -- & -- & $\phantom{1}0.37 \pm 0.07\phantom{1}$ & $\phantom{10}0.45 \pm 0.09\phantom{1}$ & $\phantom{10}0.45 \pm 0.09\phantom{1}$ & $\phantom{10}0.47 \pm 0.09\phantom{1}$ & -- \\
23 & G104.7622+02.7179 & -- & -- & -- & $\phantom{1}0.54 \pm 0.11\phantom{1}$ & $\phantom{10}0.74 \pm 0.15\phantom{1}$ & $\phantom{10}0.93 \pm 0.19\phantom{1}$ & $\phantom{10}0.88 \pm 0.18\phantom{1}$ & $\phantom{10}3.29 \pm 0.66\phantom{10}$ \\
24 & G104.7631+02.7173 & -- & -- & -- & $\phantom{1}0.49 \pm 0.10\phantom{1}$ & $\phantom{10}0.78 \pm 0.16\phantom{1}$ & $\phantom{10}1.00 \pm 0.20\phantom{1}$ & $\phantom{10}1.01 \pm 0.20\phantom{1}$ & -- \\
25 & G104.7934+02.6942 & -- & $\phantom{10}0.51 \pm 0.10\phantom{1}$ & $\phantom{10}0.82 \pm 0.16\phantom{1}$ & $\phantom{1}1.39 \pm 0.28\phantom{1}$ & $\phantom{10}1.96 \pm 0.39\phantom{1}$ & $\phantom{10}2.54 \pm 0.51\phantom{1}$ & $\phantom{10}3.25 \pm 0.65\phantom{1}$ & $\phantom{10}3.37 \pm 0.67\phantom{10}$ \\
26 & G104.8126+02.6920 & -- & -- & -- & $\phantom{1}2.74 \pm 0.55\phantom{1}$ & $\phantom{10}6.24 \pm 1.25\phantom{1}$ & $\phantom{1}11.99 \pm 2.40\phantom{1}$ & $\phantom{1}18.67 \pm 3.73\phantom{1}$ & $\phantom{1}22.97 \pm 4.59\phantom{10}$ \\
27 & G104.8206+02.7026 & -- & -- & -- & $\phantom{1}0.27 \pm 0.05\phantom{1}$ & $\phantom{10}0.44 \pm 0.09\phantom{1}$ & $\phantom{10}0.63 \pm 0.13\phantom{1}$ & $\phantom{10}0.97 \pm 0.19\phantom{1}$ & -- \\
28 & G104.8225+02.8393 & -- & -- & -- & $\phantom{1}0.66 \pm 0.13\phantom{1}$ & $\phantom{10}1.10 \pm 0.22\phantom{1}$ & $\phantom{10}1.79 \pm 0.36\phantom{1}$ & $\phantom{10}2.43 \pm 0.49\phantom{1}$ & $\phantom{10}3.53 \pm 0.71\phantom{10}$ \\
29 & G104.8383+02.7091 & -- & $\phantom{10}0.37 \pm 0.07\phantom{1}$ & $\phantom{10}1.71 \pm 0.34\phantom{1}$ & $\phantom{1}4.13 \pm 0.83\phantom{1}$ & $\phantom{10}5.74 \pm 1.15\phantom{1}$ & $\phantom{10}8.73 \pm 1.75\phantom{1}$ & $\phantom{1}10.28 \pm 2.06\phantom{1}$ & $\phantom{1}11.31 \pm 2.26\phantom{10}$ \\
30 & G104.8429+02.6522 & -- & -- & -- & $\phantom{1}0.12 \pm 0.02\phantom{1}$ & $\phantom{10}0.17 \pm 0.03\phantom{1}$ & $\phantom{10}0.28 \pm 0.06\phantom{1}$ & $\phantom{10}0.38 \pm 0.08\phantom{1}$ & -- \\
31 & G104.8436+02.7289 & -- & -- & -- & $\phantom{1}0.11 \pm 0.02\phantom{1}$ & $\phantom{10}0.17 \pm 0.03\phantom{1}$ & $\phantom{10}0.25 \pm 0.05\phantom{1}$ & $\phantom{10}0.32 \pm 0.06\phantom{1}$ & -- \\
32 & G104.8510+02.7998 & -- & -- & -- & $\phantom{1}2.21 \pm 0.44\phantom{1}$ & $\phantom{10}2.77 \pm 0.55\phantom{1}$ & $\phantom{10}3.52 \pm 0.70\phantom{1}$ & $\phantom{10}4.00 \pm 0.80\phantom{1}$ & $\phantom{10}4.58 \pm 0.92\phantom{10}$ \\
33 & G104.8965+02.7124 & -- & -- & -- & $\phantom{1}0.54 \pm 0.11\phantom{1}$ & $\phantom{10}0.50 \pm 0.10\phantom{1}$ & $\phantom{10}0.39 \pm 0.08\phantom{1}$ & $\phantom{10}0.30 \pm 0.06\phantom{1}$ & $\phantom{10}2.98 \pm 0.60\phantom{10}$ \\
34 & G104.9084+02.7625 & $\phantom{10}1.44 \pm 0.29\phantom{1}$ & $\phantom{10}2.68 \pm 0.54\phantom{1}$ & $\phantom{10}3.80 \pm 0.76\phantom{1}$ & $\phantom{1}6.81 \pm 1.36\phantom{1}$ & $\phantom{10}7.93 \pm 1.59\phantom{1}$ & $\phantom{10}8.26 \pm 1.65\phantom{1}$ & $\phantom{10}8.01 \pm 1.60\phantom{1}$ & $\phantom{10}6.62 \pm 1.32\phantom{10}$ \\
35 & G104.9180+02.7798 & $\phantom{10}0.27 \pm 0.05\phantom{1}$ & $\phantom{10}0.59 \pm 0.12\phantom{1}$ & -- & $\phantom{1}0.19 \pm 0.04\phantom{1}$ & $\phantom{10}0.12 \pm 0.03\phantom{1}$ & -- & -- & $\phantom{10}1.03 \pm 0.21\phantom{10}$ \\
\hline
\end{tabular}
\end{small}
\end{center}
\end{sidewaystable*}

\begin{sidewaystable*}[ph!]
\begin{center}
\caption{Photometry of YSO candidates in the \iras{} field from mid-infrared to submillimeter wavelengths. The columns represent the AllWISE 22\,${\rm \mu m}$, {\it Spitzer} MIPS 24\,${\rm \mu m}$ and {\it Herschel} PACS and SPIRE bands. Units are mJy and the upper limits are $3\sigma$. \label{tab:photo2}}
\setlength{\tabcolsep}{1 mm}
\begin{small}
\begin{tabular}{cccccccccc}
\hline
\hline
Nr & IRAC Designation & WISE & MIPS & \multicolumn{3}{c}{PACS} & \multicolumn{3}{c}{SPIRE} \\
~ & SSTSMOGA & 22 $\mu$m & 24 $\mu$m & 70 $\mu$m & 100 $\mu$m & 160 $\mu$m & 250 $\mu$m & 350 $\mu$m & 500 $\mu$m \\
\hline
1 & G104.5582+02.8857 & -- & $\phantom{10}3.46 \pm 0.69\phantom{10}$ & -- & -- & -- & -- & -- & -- \\
2 & G104.6313+02.6211 & -- & -- & -- & -- & -- & $<26.25$ & $<19.13$ & $<20.39$ \\
3 & G104.6697+02.9357 & $\phantom{100}4.52 \pm 1.20\phantom{10}$ & $\phantom{10}5.51 \pm 1.10\phantom{10}$ & -- & -- & -- & $<543.11$ & $<439.78$ & $<354.30$ \\
4 & G104.6829+02.6173 & -- & $\phantom{10}3.07 \pm 0.61\phantom{10}$ & -- & -- & -- & $<14.11$ & $<14.21$ & $<14.89$ \\
5 & G104.6895+02.7945 & $1441.33 \pm 288.27$ & $951.30 \pm 190.26$ & $3812.09 \pm 762.42\phantom{1}$ & $4602.46 \pm 920.49\phantom{1}$ & $\phantom{1}6556.28 \pm 1311.26$ & $4978.77 \pm 995.75\phantom{1}$ & $2566.71 \pm 513.34$ & $2656.18 \pm 531.24$ \\
6 & G104.6968+02.7925 & $\phantom{1}172.81 \pm 34.56\phantom{1}$ & -- & $<638.26$ & $<1629.24$ & $<4786.60$ & $<4657.96$ & $<4425.38$ & $<3035.44$ \\
7 & G104.7003+02.7923 & -- & -- & $\phantom{1}970.26 \pm 194.05\phantom{1}$ & $2180.21 \pm 436.04\phantom{1}$ & $\phantom{1}5872.59 \pm 1174.52$ & $5667.15 \pm 1133.43$ & $4845.84 \pm 969.17$ & $3304.47 \pm 660.89$ \\
8 & G104.7009+02.6186 & -- & $\phantom{10}0.76 \pm 0.15\phantom{10}$ & -- & -- & -- & $<17.52$ & $<11.93$ & $<16.96$ \\
9 & G104.7023+02.8142 & $<2.34$ & -- & $<137.91$ & $<264.69$ & $<698.90$ & $<546.92$ & $<416.00$ & $<383.71$ \\
10 & G104.7061+02.7861 & -- & -- & $<1205.16$ & $<2483.55$ & $<5822.72$ & $<5420.34$ & $<3622.00$ & $<3088.32$ \\
11 & G104.7062+02.7895 & $\phantom{1}652.17 \pm 130.43$ & -- & $<1266.84$ & $<2953.93$ & $<6997.07$ & $<5817.88$ & $<4658.27$ & $<3651.07$ \\
12 & G104.7118+02.7921 & $\phantom{1}837.84 \pm 167.57$ & -- & $<840.59$ & $<2140.81$ & $<5910.42$ & $<4916.79$ & $<4984.93$ & $<3561.18$ \\
13 & G104.7153+02.7979 & $\phantom{1}331.71 \pm 66.34\phantom{1}$ & -- & $<897.40$ & $<1881.94$ & $<3962.98$ & $<2977.20$ & $<2390.63$ & $<1994.98$ \\
14 & G104.7237+02.7946 & -- & -- & $<298.37$ & $<835.45$ & $<2920.31$ & $<4428.71$ & $<3691.58$ & $<2952.51$ \\
15 & G104.7241+02.8075 & $\phantom{1}124.73 \pm 24.95\phantom{1}$ & -- & $<215.06$ & $<562.18$ & $<1561.44$ & $<1580.04$ & $<1315.64$ & $<1074.96$ \\
16 & G104.7270+02.8003 & $2400.83 \pm 480.17$ & -- & $5572.69 \pm 1114.54$ & $7435.41 \pm 1487.08$ & $10568.03 \pm 2113.61$ & $5931.67 \pm 1186.33$ & $3791.80 \pm 758.36$ & $2551.67 \pm 510.33$ \\
17 & G104.7418+02.5949 & $<1.45$ & -- & -- & -- & -- & -- & $<11.62$ & -- \\
18 & G104.7513+02.5857 & $<1.74$ & -- & -- & -- & -- & -- & -- & -- \\
19 & G104.7538+02.8964 & $\phantom{100}6.25 \pm 1.25\phantom{10}$ & -- & $<68.21$ & $<168.90$ & $<529.21$ & $<425.21$ & $<317.81$ & $<269.32$ \\
20 & G104.7544+02.6101 & $\phantom{10}16.78 \pm 3.36\phantom{10}$ & $\phantom{1}14.68 \pm 2.94\phantom{10}$ & -- & -- & -- & -- & $<13.41$ & $<17.90$ \\
21 & G104.7601+02.7154 & -- & -- & $<38.26$ & $<122.08$ & $<507.21$ & $<800.02$ & $<666.55$ & $<579.15$ \\
22 & G104.7611+02.7176 & -- & -- & $<37.69$ & $<110.58$ & $<706.18$ & $<872.82$ & $<940.50$ & $<634.56$ \\
23 & G104.7622+02.7179 & $\phantom{10}21.98 \pm 4.40\phantom{10}$ & -- & $<70.86$ & $<194.06$ & $<781.75$ & $<1034.92$ & $<1113.86$ & $<634.56$ \\
24 & G104.7631+02.7173 & -- & -- & $<60.01$ & $<198.57$ & $<771.71$ & $<1034.92$ & $<1113.86$ & $<634.56$ \\
25 & G104.7934+02.6942 & $\phantom{10}13.98 \pm 2.80\phantom{10}$ & $\phantom{1}13.45 \pm 2.69\phantom{10}$ & -- & -- & -- & $<351.64$ & $<540.90$ & $<398.58$ \\
26 & G104.8126+02.6920 & $\phantom{10}54.85 \pm 10.97\phantom{1}$ & $\phantom{1}36.49 \pm 7.30\phantom{10}$ & -- & -- & -- & -- & -- & -- \\
27 & G104.8206+02.7026 & -- & $\phantom{10}2.85 \pm 0.57\phantom{10}$ & -- & -- & -- & -- & -- & -- \\
28 & G104.8225+02.8393 & $\phantom{100}9.22 \pm 1.84\phantom{10}$ & $\phantom{10}6.81 \pm 1.36\phantom{10}$ & -- & -- & -- & $<340.48$ & $<334.73$ & -- \\
29 & G104.8383+02.7091 & $\phantom{10}27.90 \pm 5.58\phantom{10}$ & $\phantom{1}11.41 \pm 2.28\phantom{10}$ & -- & -- & -- & -- & -- & -- \\
30 & G104.8429+02.6522 & -- & $\phantom{10}0.64 \pm 0.13\phantom{10}$ & -- & -- & -- & -- & -- & -- \\
31 & G104.8436+02.7289 & -- & $\phantom{10}1.10 \pm 0.22\phantom{10}$ & -- & -- & -- & -- & -- & -- \\
32 & G104.8510+02.7998 & $\phantom{10}20.34 \pm 4.07\phantom{10}$ & $\phantom{1}12.58 \pm 2.52\phantom{10}$ & -- & -- & -- & -- & -- & -- \\
33 & G104.8965+02.7124 & $\phantom{100}6.85 \pm 1.37\phantom{10}$ & $\phantom{10}0.65 \pm 0.13\phantom{10}$ & -- & -- & -- & -- & -- & -- \\
34 & G104.9084+02.7625 & $\phantom{100}6.20 \pm 1.24\phantom{10}$ & $\phantom{10}4.54 \pm 0.91\phantom{10}$ & -- & -- & -- & -- & -- & -- \\
35 & G104.9180+02.7798 & $\phantom{100}4.27 \pm 0.86\phantom{10}$ & -- & -- & -- & -- & -- & -- & -- \\
\hline
\end{tabular}
\end{small}
\end{center}
\end{sidewaystable*}

\begin{sidewaystable*}[ph!]
\begin{center}
\caption{Multi-wavelength photometry of YSO candidates in the \iras{} field with unknown distances. The columns represent the 2MASS $JHK_s$, {\it Spitzer} IRAC, AllWISE 12 and 22\,${\rm \mu m}$ and {\it Spitzer} MIPS 24\,${\rm \mu m}$. Units are in mJy. \label{tab:photo_no_d}}
\setlength{\tabcolsep}{2 mm}
\begin{small}
\begin{tabular}{cccccccccccc}
\hline
\hline
Nr & IRAC Designation & \multicolumn{3}{c}{2MASS} & \multicolumn{4}{c}{IRAC} & \multicolumn{2}{c}{WISE} & MIPS \\
~ & SSTSMOGA & $J$ & $H$ & $K_s$ & 3.6 $\mu$m & 4.5 $\mu$m & 5.8 $\mu$m & 8.0 $\mu$m & 12 $\mu$m & 22 $\mu$m & 24 $\mu$m \\
\hline
1 & G104.5742+02.6247 & $0.36 \pm 0.06$ & $0.54 \pm 0.07$ & $0.82 \pm 0.09$ & $\phantom{1}0.77 \pm 0.03$ & $\phantom{1}0.80 \pm 0.03$ & $\phantom{1}0.75 \pm 0.05$ & $\phantom{1}0.72 \pm 0.04$ & $\phantom{10}2.13 \pm 0.15$ & $\phantom{100}5.92 \pm 0.84\phantom{1}$ & $0.90 \pm 0.26$ \\
2 & G104.5824+02.6207 & -- & -- & -- & $\phantom{1}0.46 \pm 0.03$ & $\phantom{1}0.77 \pm 0.03$ & $\phantom{1}1.07 \pm 0.05$ & $\phantom{1}1.52 \pm 0.07$ & $\phantom{10}3.96 \pm 0.15$ & $\phantom{10}15.69 \pm 1.14\phantom{1}$ & $3.83 \pm 0.37$ \\
3 & G104.5939+02.8413 & $0.57 \pm 0.05$ & $0.55 \pm 0.07$ & $0.53 \pm 0.09$ & $\phantom{1}0.24 \pm 0.03$ & $\phantom{1}0.16 \pm 0.01$ & -- & -- & $\phantom{10}1.50 \pm 0.10$ & $\phantom{100}4.56 \pm 0.86\phantom{1}$ & -- \\
4 & G104.5956+02.8382 & -- & -- & -- & $\phantom{1}0.14 \pm 0.03$ & $\phantom{1}0.11 \pm 0.01$ & -- & -- & $\phantom{10}1.55 \pm 0.12$ & $\phantom{100}4.38 \pm 0.80\phantom{1}$ & -- \\
5 & G104.5984+02.7886 & -- & -- & -- & $\phantom{1}0.10 \pm 0.01$ & $\phantom{1}0.09 \pm 0.01$ & -- & -- & $\phantom{10}0.95 \pm 0.13$ & $\phantom{100}3.76 \pm 0.81\phantom{1}$ & -- \\
6 & G104.6468+02.8399 & -- & -- & -- & $\phantom{1}0.28 \pm 0.01$ & $\phantom{1}0.20 \pm 0.01$ & -- & -- & $\phantom{10}2.50 \pm 0.21$ & $\phantom{100}9.81 \pm 1.13\phantom{1}$ & -- \\
7 & G104.6796+02.8554 & -- & -- & -- & $10.47 \pm 0.46$ & $31.65 \pm 0.79$ & $63.56 \pm 1.32$ & $78.03 \pm 1.71$ & $285.06 \pm 3.94$ & $1054.61 \pm 22.34$ & -- \\
8 & G104.6993+02.7714 & $3.93 \pm 0.09$ & $6.65 \pm 0.07$ & $6.19 \pm 0.09$ & $\phantom{1}3.57 \pm 0.10$ & $\phantom{1}2.29 \pm 0.06$ & $\phantom{1}1.63 \pm 0.09$ & -- & $\phantom{1}22.66 \pm 0.48$ & $\phantom{1}237.19 \pm 7.43\phantom{1}$ & -- \\
9 & G104.7034+02.7674 & $3.19 \pm 0.10$ & $4.06 \pm 0.07$ & $3.15 \pm 0.09$ & $\phantom{1}1.62 \pm 0.05$ & $\phantom{1}1.04 \pm 0.03$ & $\phantom{1}0.65 \pm 0.08$ & -- & $\phantom{1}25.76 \pm 0.44$ & $\phantom{10}92.02 \pm 4.32\phantom{1}$ & -- \\
10 & G104.7076+02.7809 & $1.50 \pm 0.05$ & $1.80 \pm 0.09$ & $1.41 \pm 0.09$ & $\phantom{1}0.73 \pm 0.04$ & $\phantom{1}0.47 \pm 0.03$ & -- & -- & $\phantom{1}87.69 \pm 1.94$ & $\phantom{1}658.70 \pm 16.99$ & -- \\
11 & G104.7301+02.7741 & -- & -- & -- & $\phantom{1}0.23 \pm 0.03$ & $\phantom{1}0.32 \pm 0.02$ & -- & -- & $\phantom{1}11.40 \pm 0.38$ & $\phantom{10}42.77 \pm 2.21\phantom{1}$ & -- \\
12 & G104.8880+02.6685 & -- & -- & -- & $\phantom{1}0.33 \pm 0.01$ & $\phantom{1}0.33 \pm 0.01$ & $\phantom{1}0.38 \pm 0.03$ & $\phantom{1}1.78 \pm 0.06$ & $\phantom{10}1.54 \pm 0.12$ & $\phantom{100}8.22 \pm 0.95\phantom{1}$ & $8.31 \pm 0.23$ \\
13 & G104.9019+02.7100 & -- & -- & -- & $\phantom{1}0.33 \pm 0.02$ & $\phantom{1}0.43 \pm 0.01$ & $\phantom{1}0.61 \pm 0.04$ & $\phantom{1}1.15 \pm 0.05$ & $\phantom{10}1.97 \pm 0.24$ & $\phantom{100}5.51 \pm 0.83\phantom{1}$ & $4.22 \pm 0.45$ \\
14 & G104.9468+02.8214 & -- & -- & -- & $\phantom{1}0.19 \pm 0.01$ & $\phantom{1}0.12 \pm 0.01$ & -- & -- & $\phantom{10}0.56 \pm 0.09$ & $\phantom{100}2.28 \pm 0.67\phantom{1}$ & -- \\
\hline
\end{tabular}
\end{small}
\end{center}
\end{sidewaystable*}

\begin{table*} 
\caption{YSO classification obtained in this work and in \citet{Winston2019}} \label{t:winston}      
\centering                        
\begin{tabular}{rcccccccccc}
\hline \hline                 
Nr & IRAC Designation\tablefootmark{a} & \multicolumn{3}{c}{Class (this work)\tablefootmark{b}}  & ~ & \multicolumn{2}{c}{Class \citep{Winston2019}\tablefootmark{c}}\\ \cline{3-5} \cline{7-8} 
~ & `SSTSMOGA'  & ML & CC & CC (no dist.) & ~ & IRAC & MIPS \\ 
\hline
1 & G104.5582+02.8857 & II$^{(t)}$\tablefootmark{d} & II & II & ~ & II & II \\
2 & G104.5742+02.6247 & II & -- & II & ~ & II & II \\
3 & G104.5824+02.6207 & I & -- & I & ~ &  I & I \\
4 & G104.6796+02.8554 & I & -- & I & ~ &  I & I \\
5 & G104.6829+02.6173 & II$^{(t)}$ & II & II & ~ &  II & II \\
6 & G104.6895+02.7945 & I$^{(t)}$ & I & I & ~ &  I & I \\
7 & \textbf{G104.6968+02.7925} & II & -- & -- & ~ &  I & I \\  
8 & G104.7003+02.7923 & I$^{(t)}$ & I & I & ~ &  I & I \\
9 & G104.7023+02.8142 & II$^{(t)}$ & II & II & ~ &  II & II \\
10 & G104.7237+02.7946 & II$^{(t)}$ & II & II & ~ &  II & II \\
11 & G104.7270+02.8003 & I$^{(t)}$ & I & I & ~ &  I & I \\
12 & G104.7513+02.5857 & II$^{(t)}$ & II & II & ~ &  II & II \\
13 & \textbf{G104.7544+02.6101} & I$^{(t)}$ & I & I & ~ &  II & II \\
14 & G104.7601+02.7154 & II$^{(t)}$ & II & II & ~ &  II & II \\
15 & G104.7611+02.7176 & II$^{(t)}$ & II & II & ~ &  II & II \\
16 & \textbf{G104.7622+02.7179} &  I$^{(t)}$ & I & I & ~ &  II & II \\
17 & \textbf{G104.7631+02.7173} &  I$^{(t)}$ & I & I & ~ &  II & II \\
18 & G104.7934+02.6942 & I$^{(t)}$& I & I & ~ &  I & I \\
19 & G104.8126+02.6920 & I$^{(t)}$& I & I & ~ &  I & I \\
20 & G104.8206+02.7026 & I$^{(t)}$ & I & I & ~ &  I & I \\
21 & G104.8225+02.8393 & I$^{(t)}$ & I & I & ~ &  I & I \\
22 & \textbf{G104.8383+02.7091} & II$^{(t)}$ & II & I & ~ &  I & II \\
23 & \textbf{G104.8510+02.7998} & I$^{(t)}$ & I & I & ~ &  II & II \\
24 & \textbf{G104.9019+02.7100} & II & -- & I & ~ &  I & I \\ 
\hline 
\end{tabular}
\begin{flushleft}
\tablefoot{
\tablefoottext{a}{For sources with different classifications obtained here and in \cite{Winston2019}, the names are written in bold.}
\tablefoottext{b}{ML, CC, and CC (no dist.) refer to the classifications obtained using machine learning (Section \ref{sec:ml}), color-color diagrams when the kinematic distance was also available (Section \ref{sec:cc}), and color-color diagrams created using the criteria which did not require distance (see text).}
\tablefoottext{c}{IRAC and MIPS refer to the classification obtained  in \cite{Winston2019} using only IRAC bands, and IRAC and MIPS bands, respectively.}
\tablefoottext{d}{For sources which belonged to the training or test samples in the ML calculations, the class is adopted from the CC method and marked with $^{(t)}$.}
}
\end{flushleft}
\end{table*}

\section{Bolometric luminosities of YSOs}
\label{sec:lbol}

We used the photometry of YSOs in the IRAS 22147 region to calculate bolometric luminositeis, $L_\mathrm{bol}$, using trapezoidal summation (Table \ref{tab:photo} and \ref{tab:photo2}). For sources with available \textit{Spitzer}/MIPS photometry at 24 $\mu$m, we omit the WISE 12 and 22 $\mu$m points due to the larger beam and possible source confusion. In the calculations, we used upper limits to constrain the SED shape in the far-IR range (see Appendix \ref{sec:seds2}) and, thus, the calculated luminosities shall be considered as upper limits of the actual $L_\mathrm{bol}$.

\begin{figure}[ht!]
\centering
\includegraphics[scale=0.8]{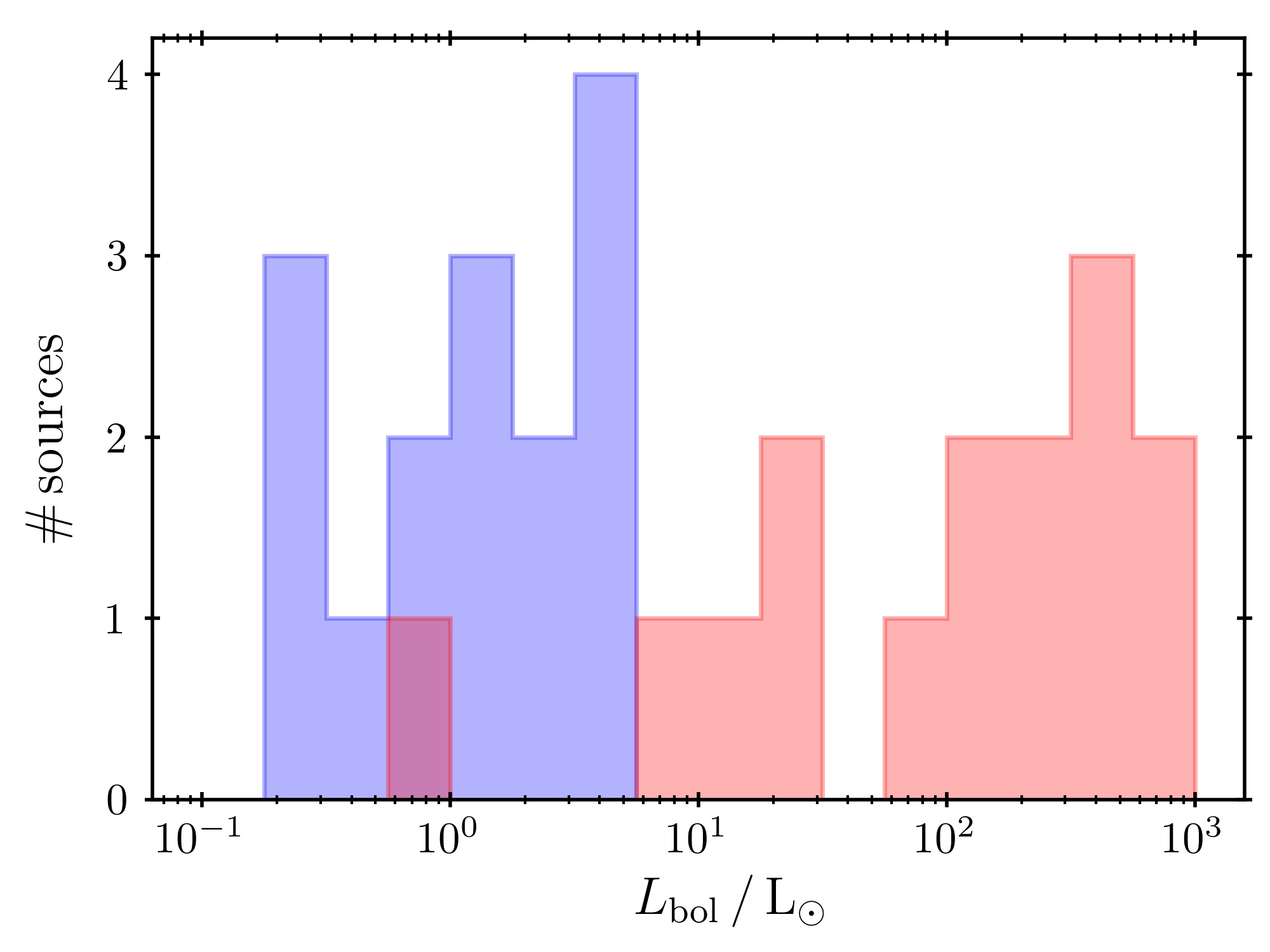}
\caption{Histogram of bolometric luminosities for YSOs located at $\sim$2.2 kpc (in blue) and at $\sim$5.6 kpc (in red), adopting kinematic distances.}
\label{hist_lbol}
\end{figure}

Figure \ref{hist_lbol} shows the distribution of the resulting $L_\mathrm{bol}$ of YSOs in the IRAS22147 region. The objects are divided by their kinematic distances into two groups to illustrate the range of bolometric luminosities, which can be probed using the available infrared photometry. YSOs located at $\sim$5.6 kpc are relatively luminous objects with $L_\mathrm{bol}$ in the range of $\sim10-1000$ L$_{\odot}$, corresponding to low- and intermediate-mass YSOs. Among more nearby sources, less luminous sources are also detected in the infrared, with  $L_\mathrm{bol}$ in the range of $\sim0.1-10$ L$_{\odot}$. Note that the single source with $L_\mathrm{bol}$ below 1 L$\mathrm{\odot}$ at 5.6 kpc is G104.8206; unfortunately, its distance cannot be verified with Gaia (Appendix \ref{sec:gaia}).

\section{Multi-wavelength images of the IRAS 22147 region}
\label{sec:stamps}

We present multi-wavelength images of the IRAS 22147 region: 2MASS $J$, $H$, and $K_s$; {\it Spitzer}/SMOG IRAC 3.6, 4.5, 5.8, and 8.0 $\mu$m; {\it Spitzer}/SMOG MIPS 24 $\mu$m; AllWISE 12 and 22 $\mu$m;  {\it Herschel} PACS 70, 100, and 160 $\mu$m; and {\it Herschel} SPIRE 250, 350, and 500 $\mu$m and VLA 1.4\,GHz. All the images show the same field of view. The wavelengths and angular resolutions of the surveys are indicated in the images.

\begin{figure*}
\begin{center} 
\includegraphics[scale=0.67, trim=0cm 1cm 0cm 1cm,clip]{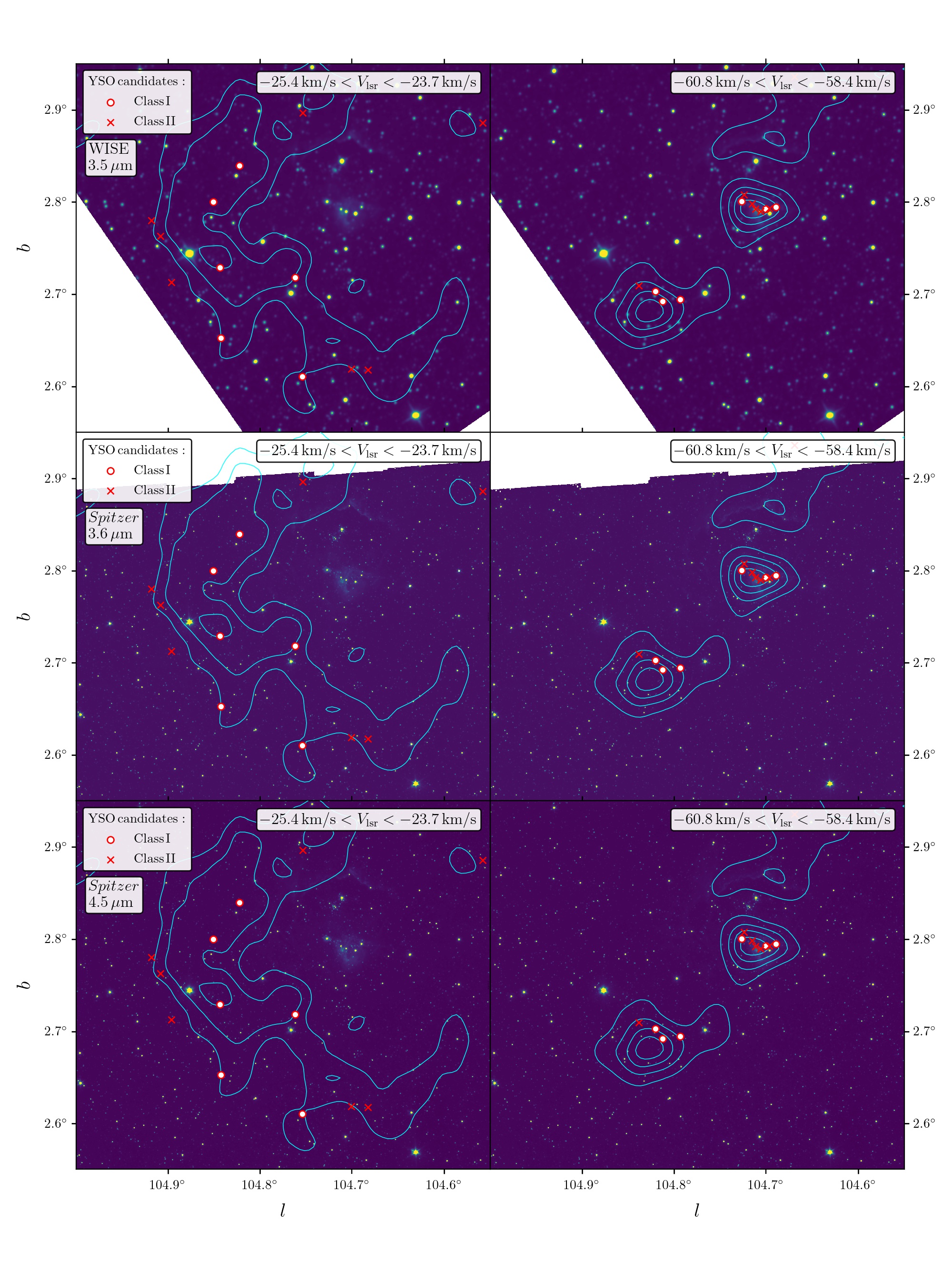} 
\end{center} \caption{Available multi-band imaging of the IRAS 22147 region, as noted in the top left corner of each row. The left and right columns in each row correspond to the two main local standard of rest (LSR) velocity ranges found in the $^{12}$CO $J=1-0$ cubes, as specified in the top right corner of each stamp. We overlaid the YSO candidates in red and the $^{12}$CO $J=1-0$ contours in cyan color. } 
\label{fig:st} 
\end{figure*}

\addtocounter{figure}{-1} \begin{figure*} \begin{center} \includegraphics[scale=0.67, trim=0cm 1cm 0cm 1cm, clip]{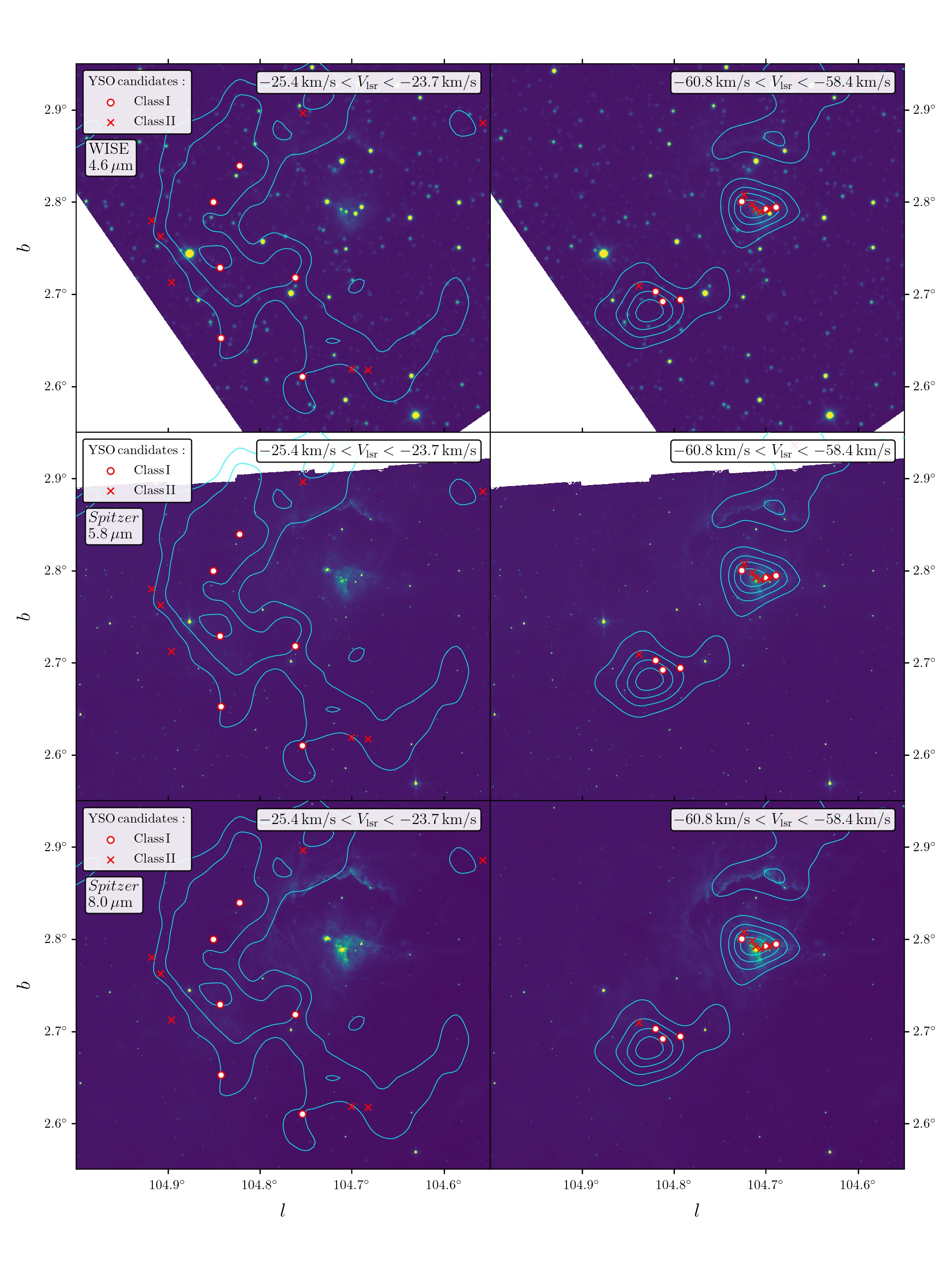} \end{center} \caption{(continued).} \end{figure*}

\addtocounter{figure}{-1} \begin{figure*} \begin{center} \includegraphics[scale=0.67, trim=0cm 1cm 0cm 1cm, clip]{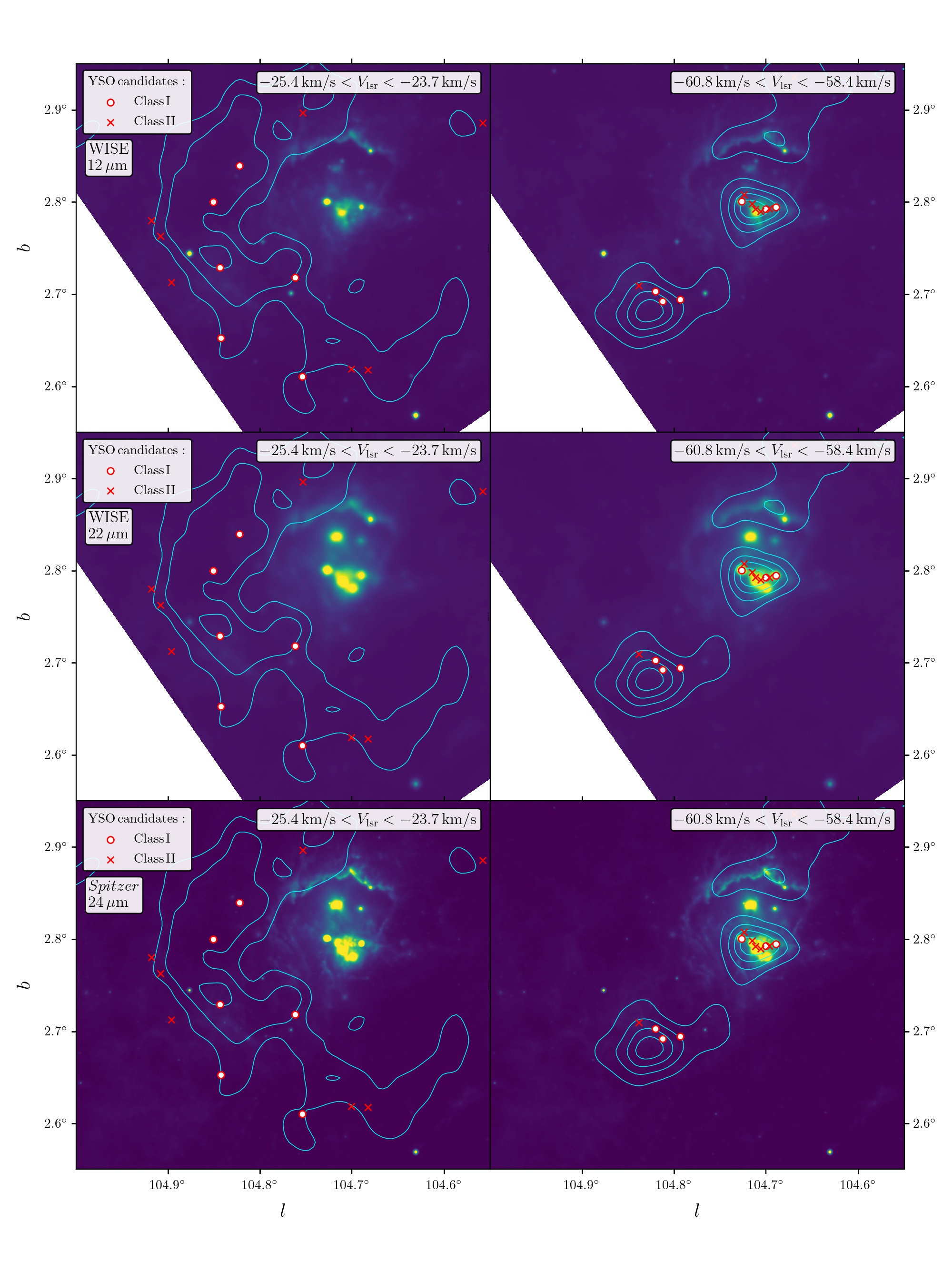} \end{center} \caption{(continued).} \end{figure*}

\addtocounter{figure}{-1} \begin{figure*} \begin{center} \includegraphics[scale=0.67, trim=0cm 1cm 0cm 1cm, clip]{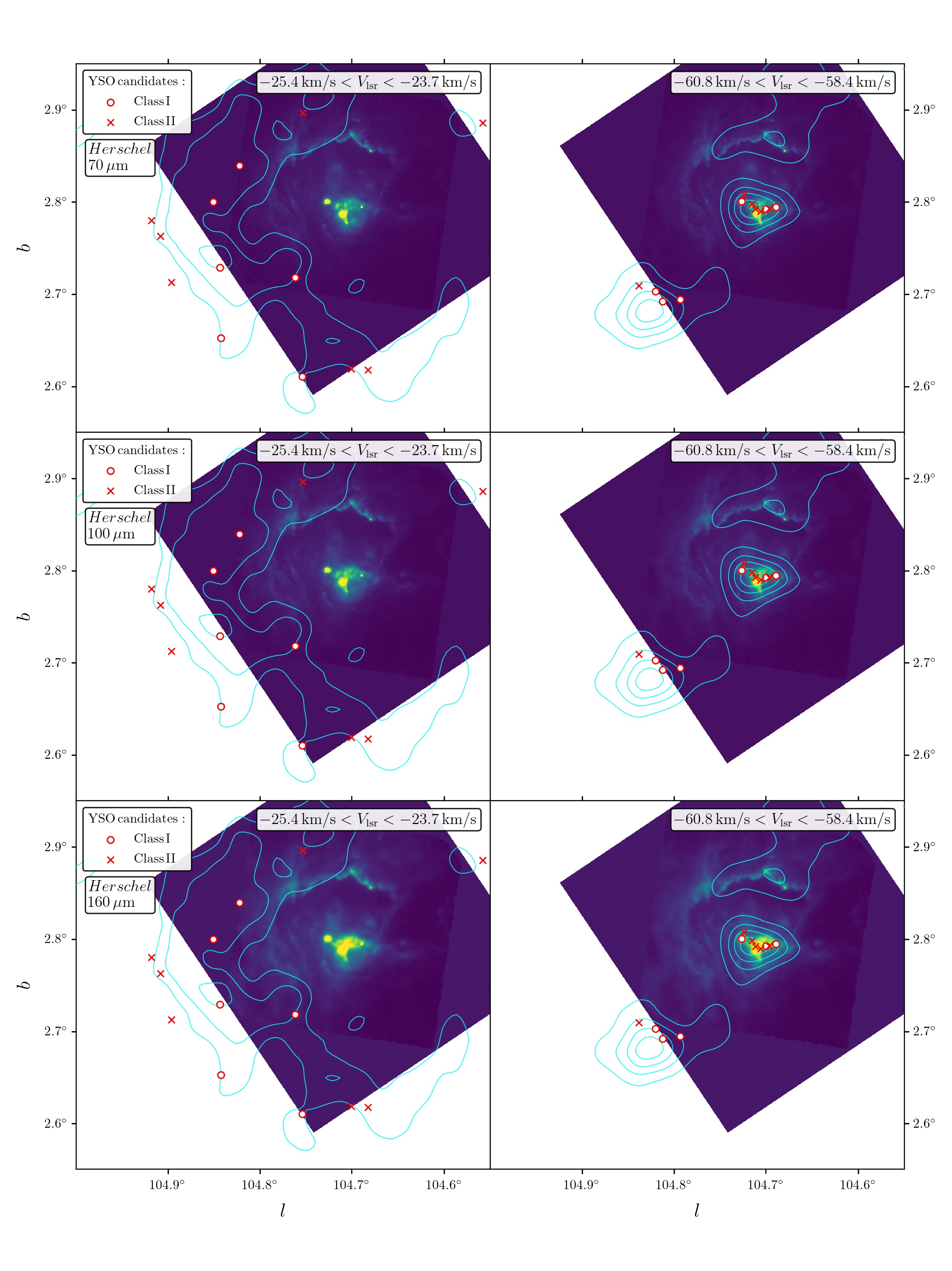} \end{center} \caption{(continued).} \end{figure*}

\addtocounter{figure}{-1} \begin{figure*} \begin{center} \includegraphics[scale=0.67, trim=0cm 1cm 0cm 1cm, clip]{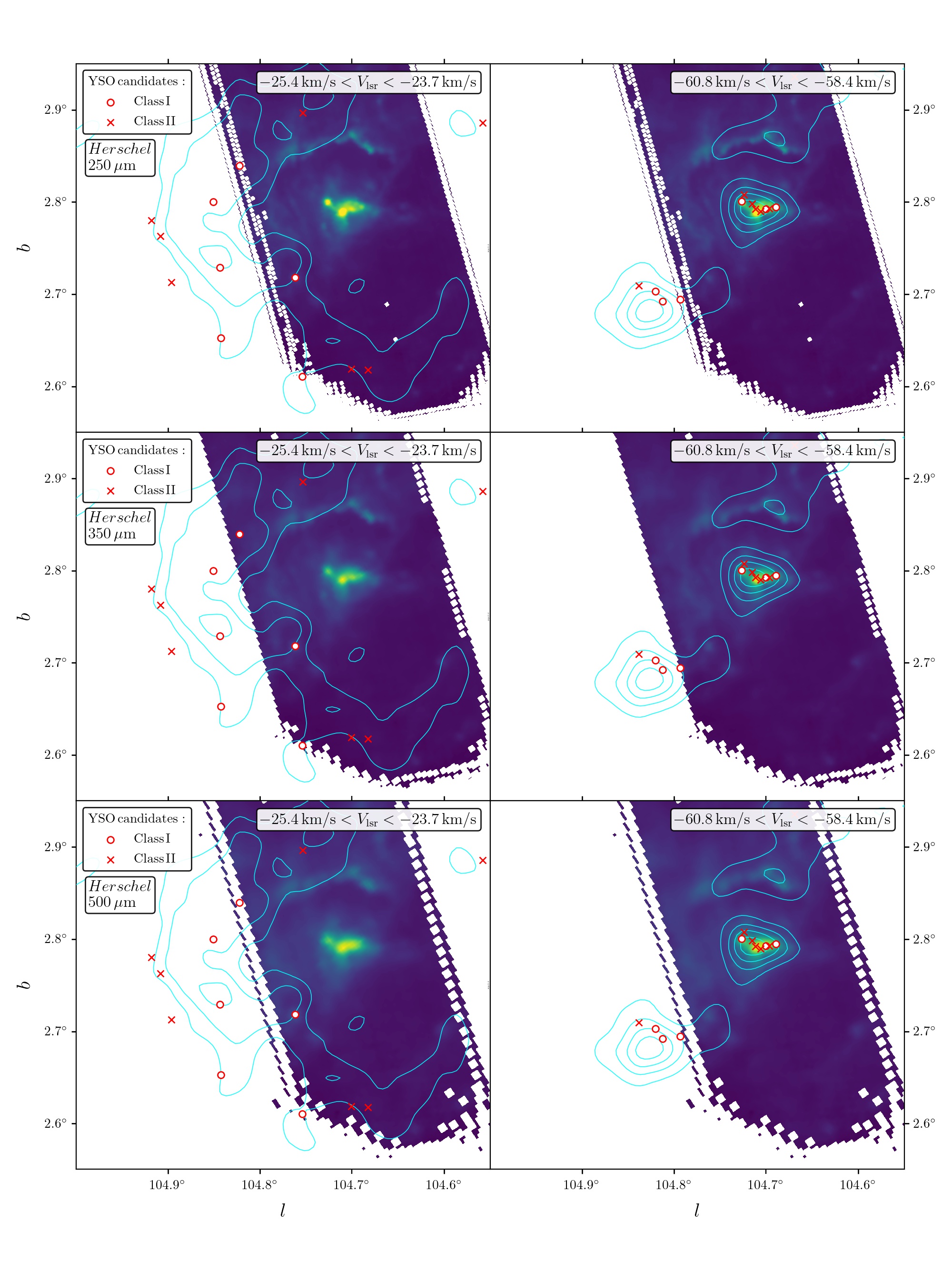} \end{center} \caption{(continued).} \end{figure*}

\addtocounter{figure}{-1} \begin{figure*} \begin{center} \includegraphics[scale=0.67, trim=0cm 1cm 0cm 1cm, clip]{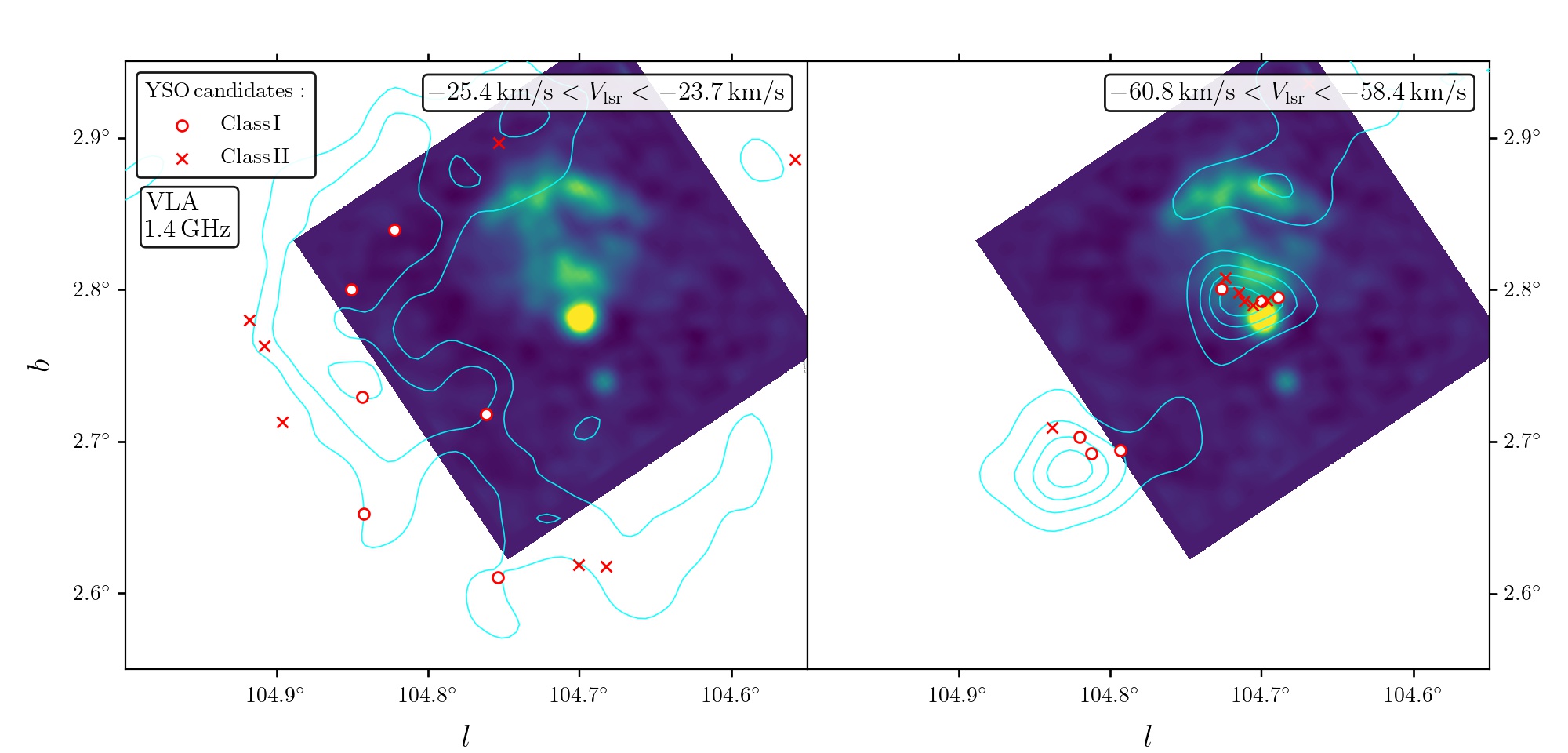} \end{center} \caption{(continued).} \end{figure*}

\section{SED fits of YSO candidates in the IRAS 22147 region}
\label{sec:seds2}

Figure \ref{fig:sed} shows the SEDs of YSOs in the IRAS 22147 region with the best-fit  \cite{robitaille2017} models. The resulting physical parameters are provided in Table \ref{t:physpar}.

\begin{figure*}
\begin{center}
\includegraphics[scale=1, trim=6cm 0cm 6cm 0cm, clip]{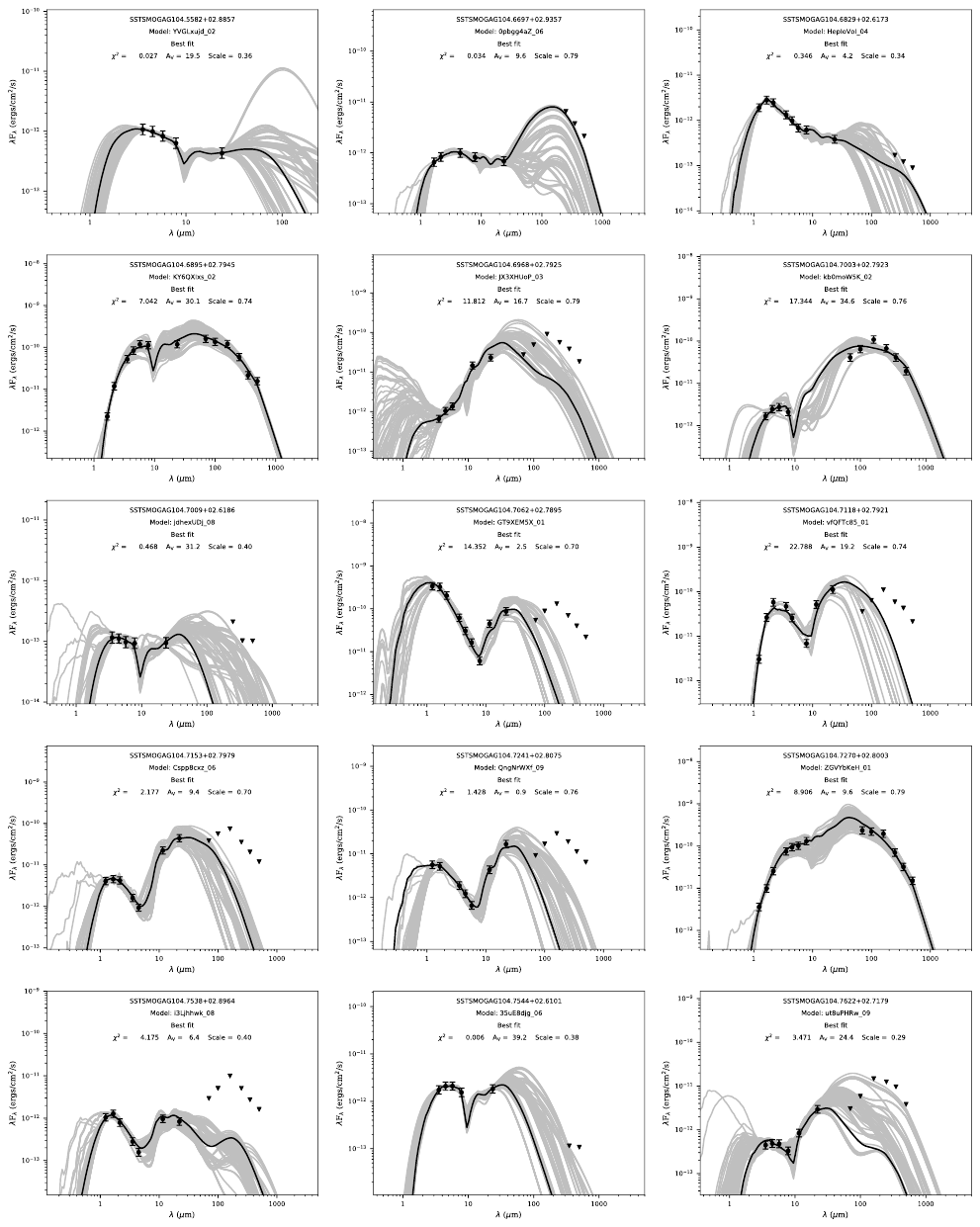}
\end{center}
\caption{SEDs of candidate YSOs with well-fit \cite{robitaille2017} YSO models. The best-fit model is indicated with the black solid line; gray lines show the YSO models with $\chi^{2}$ between $\chi^{2}_\mathrm{best}$ and $\chi^{2}_\mathrm{best}$+$F$, where $F$ is a threshold parameter which we set to 3 \citep{sewilo2019}. Filled circles and triangles are valid flux values and flux upper limits, respectively. The values of a reduced $\chi^{2}$ and interstellar visual extinction for the best-fit model are indicated in the plots.}
\label{fig:sed}
\end{figure*}

\addtocounter{figure}{-1}
\begin{figure*}
\begin{center}
\includegraphics[scale=1, trim=6cm 4cm 6cm 0cm, clip]{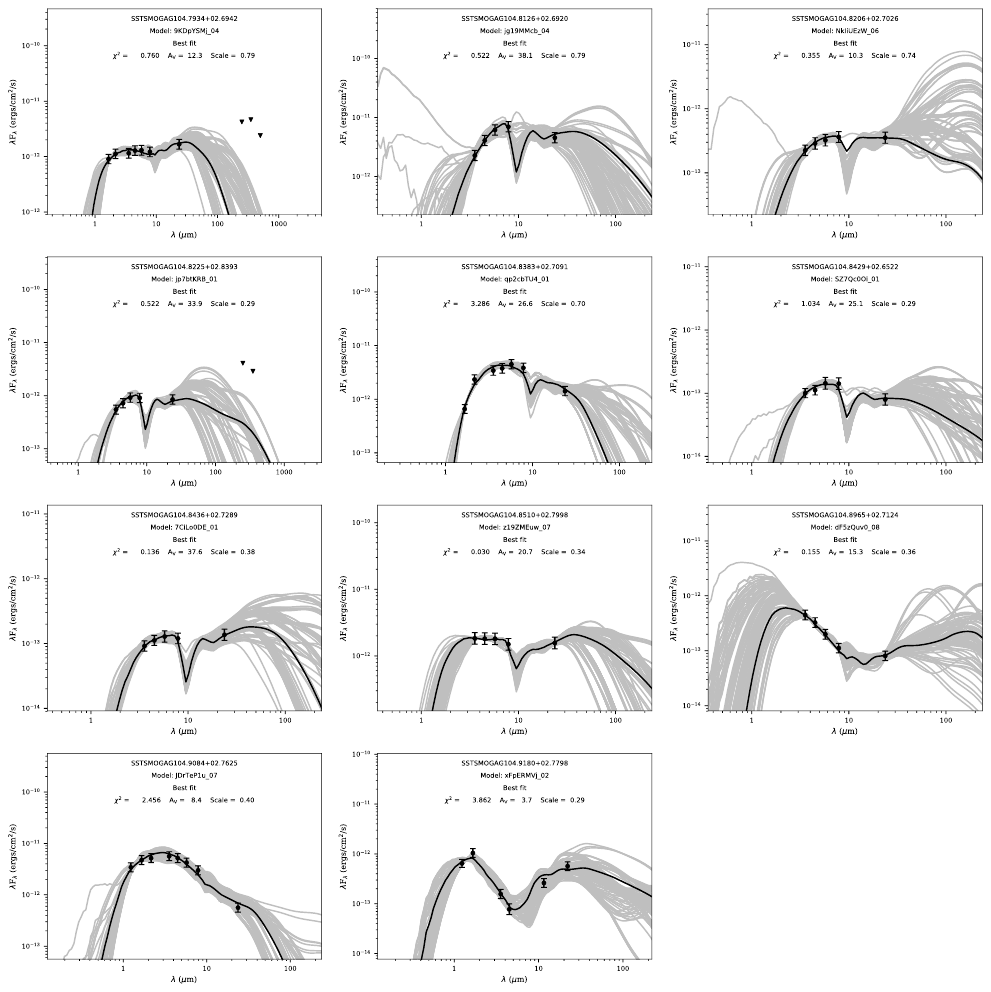}
\end{center}
\caption{(continued).}
\end{figure*}

\end{appendix}
\end{document}